\documentclass[aps,prx,twocolumn,superscriptaddress]{revtex4-2}  
\usepackage{bm,graphicx,mathrsfs}
\usepackage{graphicx}
\usepackage{epsfig}
\usepackage{amsmath,bbm}
\usepackage{amsfonts,amssymb}
\usepackage{times}
\usepackage{lipsum}

\usepackage{stmaryrd}
\usepackage{latexsym}
\usepackage{amssymb}
\usepackage{amsmath}
\usepackage{booktabs}
\usepackage[colorlinks=True
            pdfborder={0 0 0},
            linkcolor=blue,,
            allcolors=blue]{hyperref}

\usepackage{color}
\usepackage{xcolor}
\usepackage{verbatim}
\usepackage{subfig}
\usepackage{float}

\usepackage{caption}

\newcommand{\ket}[1]{|#1\rangle}

\usepackage{nameref} 
\makeatletter
\newcommand{\labelname}[1]{
  \def\@currentlabelname{#1}}%
\makeatother
\newcommand{\beginsupplement}{%
        \setcounter{table}{0}
        \renewcommand{\thetable}{S\arabic{table}}%
        \setcounter{figure}{0}
        \renewcommand{\thefigure}{S\arabic{figure}}%
     }

\begin{document}
\title{Matching and maximum likelihood decoding of a multi-round \\
subsystem quantum error correction experiment}
\author{Neereja Sundaresan}
\altaffiliation{Contributing author}
\email{neereja@ibm.com}
\affiliation{IBM Quantum, IBM T.J. Watson Research Center, Yorktown Heights, NY 10598, USA}
\author{Theodore J. Yoder}
\altaffiliation{Contributing author}
\email{ted.yoder@ibm.com}
\affiliation{IBM Quantum, IBM T.J. Watson Research Center, Yorktown Heights, NY 10598, USA}
\author{Youngseok Kim}
\affiliation{IBM Quantum, IBM T.J. Watson Research Center, Yorktown Heights, NY 10598, USA}
\author{Muyuan Li}
\affiliation{IBM Quantum, IBM T.J. Watson Research Center, Yorktown Heights, NY 10598, USA}
\author{Edward H. Chen}
\affiliation{IBM Quantum, IBM Almaden Research Center, San Jose, CA 95120, USA}
\author{Grace Harper}
\affiliation{IBM Quantum, IBM T.J. Watson Research Center, Yorktown Heights, NY 10598, USA}
\author{Ted Thorbeck}
\affiliation{IBM Quantum, IBM T.J. Watson Research Center, Yorktown Heights, NY 10598, USA}
\author{Andrew W. Cross}
\affiliation{IBM Quantum, IBM T.J. Watson Research Center, Yorktown Heights, NY 10598, USA}
\author{Antonio D. C\'orcoles}
\affiliation{IBM Quantum, IBM T.J. Watson Research Center, Yorktown Heights, NY 10598, USA}
\author{Maika Takita}
\affiliation{IBM Quantum, IBM T.J. Watson Research Center, Yorktown Heights, NY 10598, USA}

\date{\today}

\begin{abstract}
Quantum error correction offers a promising path for performing quantum computations with low errors. Although a fully fault-tolerant execution of a quantum algorithm remains unrealized, recent experimental developments, along with improvements in control electronics, are enabling increasingly advanced demonstrations of the necessary operations for applying quantum error correction. Here, we perform quantum error correction on superconducting qubits connected in a heavy-hexagon lattice. The full processor can encode a logical qubit with distance three and perform several rounds of fault-tolerant syndrome measurements that allow the correction of any single fault in the circuitry. Furthermore, by using dynamic circuits and classical computation as part of our syndrome extraction protocols, we can exploit real-time feedback to reduce the impact of energy relaxation error in the syndrome and flag qubits. We show that the logical error varies depending on the use of a perfect matching decoder compared to a maximum likelihood decoder. We observe a logical error per syndrome measurement round as low as $\sim0.04$ for the matching decoder and as low as $\sim0.035$ for the maximum likelihood decoder. Our results suggest that more significant improvements to decoders are likely on the horizon as quantum hardware has reached a new stage of development towards fully fault-tolerant operations.

\end{abstract}

\maketitle

\section{Introduction}
The outcomes of quantum computations can be faulty, in practice, due to noise in the hardware. To eliminate the resulting faults, quantum error correction (QEC) codes can be used to encode the quantum information into protected, logical degrees of freedom, and then by correcting the faults faster than they accumulate enable fault-tolerant (FT) computations. A complete execution of QEC will likely require: preparation of logical states; realization of a universal set of logical gates, which may require the preparation of magic states; repeated measurements of syndromes; and the decoding of the syndromes for correcting errors. If successful, the resulting logical error rates should be less than the underlying physical error rates, and decrease with increasing code distances down to negligible values. 

The choice of a quantum error correcting code will require consideration of the underlying hardware and its noise properties. Specifically for a heavy-hexagon lattice ~\cite{chamberlandTopologicalSubsystemCodes2020,hertzberg2021laser} of qubits, subsystem QEC codes~\cite{Poulin2005Subsystem} are attractive because they are well-suited for qubits with reduced connectivities. Other codes have shown promise due to their relatively high threshold for FT~\cite{Dennis2002TopologicalQM} or large number of transversal logical gates~\cite{Bombin06Color}. Although their space and time overhead may pose a significant hurdle for scalability, there exist encouraging approaches to reduce the most expensive resources by exploiting some form of error mitigation~\cite{Piveteau21}.

In the decoding process, successful correction depends not only on the performance of the quantum hardware, but also on the implementation of the control electronics used for acquiring and processing the classical information obtained from syndrome measurements. In our case, initializing both syndrome and flag qubits via real-time feedback between measurement cycles can help mitigate errors. At the decoding level, whereas some protocols exist to perform quantum error correction asynchronously within a FT formalism~\cite{Chamberland2018faulttolerant, DiVincenzo07}, the rate at which the error syndromes are received should be commensurate with their classical processing time to avoid an increasing backlog of syndrome data. Also, the efficient performance of some particular protocols, like using a magic state for a logical $T$-gate, require the application of real-time feed-forward. 

Thus, the long term vision of quantum error correction does not gravitate around a single ultimate goal but should be seen as a continuum of deeply interrelated tasks. The experimental path in the development of this technology will comprise the demonstration of these tasks in isolation first and their progressive combination later, always while continuously improving their associated metrics. Some of this progress is reflected in numerous recent advances on quantum systems across different physical platforms, which have demonstrated or approximated several aspects of the desiderata for FT quantum computing. In particular, FT logical state preparation has been demonstrated on ions~\cite{linkeFaulttolerantQuantumError}, nuclear spins in diamond~\cite{abobeihFaulttolerantOperationLogical2021} and superconducting qubits~\cite{takitaExperimentalDemonstrationFaulttolerant2017}. Repeated cycles of syndrome extraction have been shown in superconducting qubits in small error detecting codes~\cite{andersenRepeatedQuantumError2020a,chenExponentialSuppressionBit2021}, including partial error correction~\cite{chen2021calibrated} as well as a universal (albeit not FT) set of single-qubit gates~\cite{marquesLogicalqubitOperationsErrordetecting2021}. A FT demonstration of a universal gate set on two logical qubits has recently been reported in ions~\cite{postler2021demonstration}. In realm of error correction, there have been recent realizations of the distance-3 surface code on superconducting qubits with decoding~\cite{Krinner2021RealizingRQ} and post-selection~\cite{Zhao2021RealizingAE}, as well as a FT implementation of a dynamically protected quantum memory using the color code~\cite{ryan-andersonRealizationRealtimeFaulttolerant2021} and the FT state preparation, operation, and measurement, including its stabilizers, of a logical state in the Bacon-Shor code in ions~\cite{eganFaulttolerantControlErrorcorrected2021b,ryan-andersonRealizationRealtimeFaulttolerant2021}.

In this work we combine the capability of real-time feedback on a superconducting qubit system with a maximum likelihood decoding protocol hitherto unexplored experimentally in order to improve the survivability of logical states. We demonstrate these tools as part of the FT operation of a subsystem code~\cite{Bacon2006OperatorQE}, the heavy-hexagon code~\cite{chamberlandTopologicalSubsystemCodes2020}, on a superconducting quantum processor. This code benefits from the use of flag qubits at small cost in terms of circuit depth. By conditionally resetting each flag and syndrome qubit after each syndrome measurement cycle, we protect our d=3 system against errors arising from the noise asymmetry inherent to energy relaxation. We further exploit some recently described decoding strategies~\cite{chen2021calibrated} and extend the decoding ideas to include maximum likelihood concepts~\cite{Dennis2002TopologicalQM,pryadko2020maximum, bravyiEfficientAlgorithmsMaximum2014}.

\section{The heavy-hexagon code and multi-round circuits}

\begin{figure*}[t]
	\centering
	\subfloat[\centering]{{\includegraphics[width=0.32\textwidth]{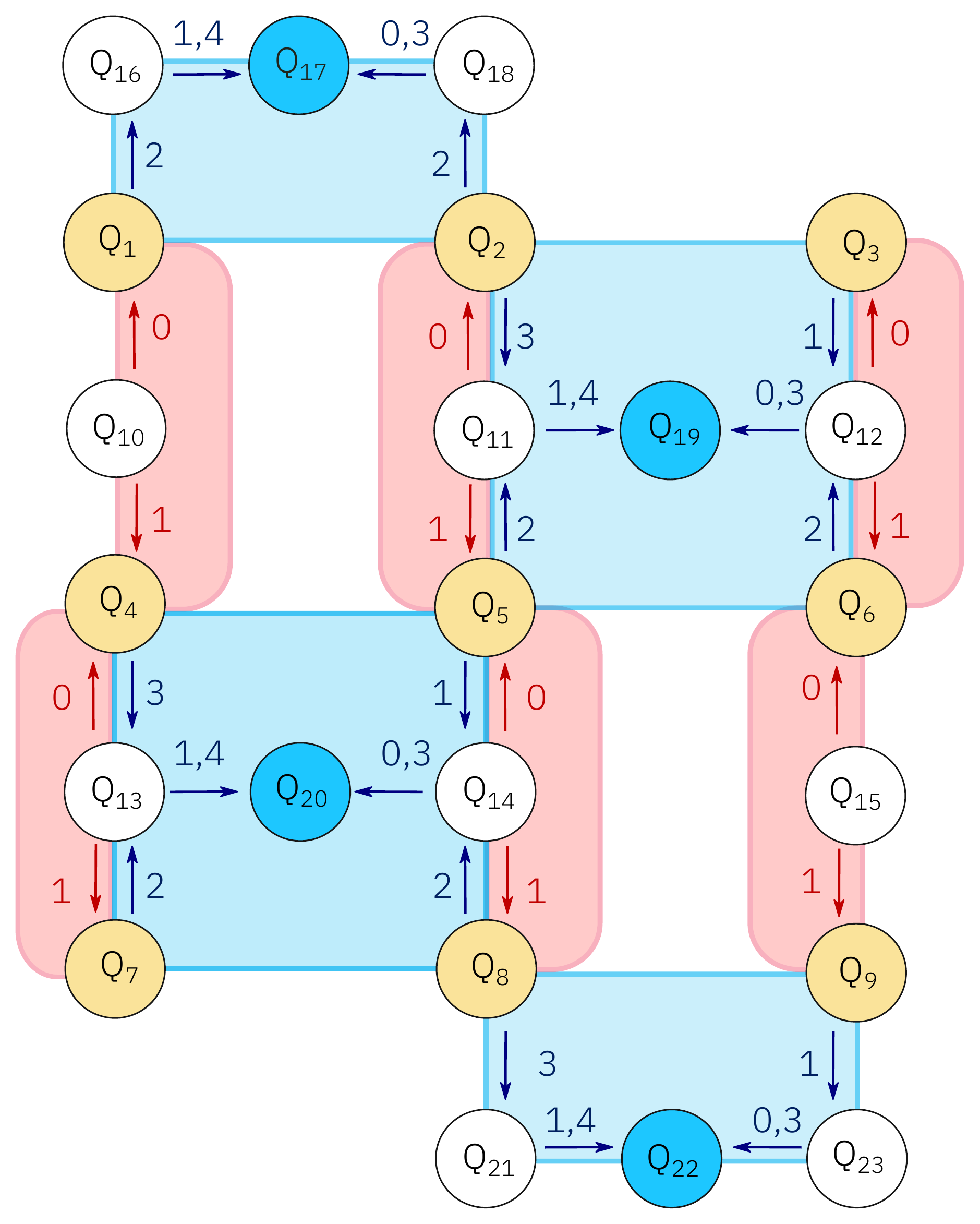}\label{fig:RoundSched_a}}}
	\qquad
	\subfloat[\centering]{{\includegraphics[width=0.64\textwidth]{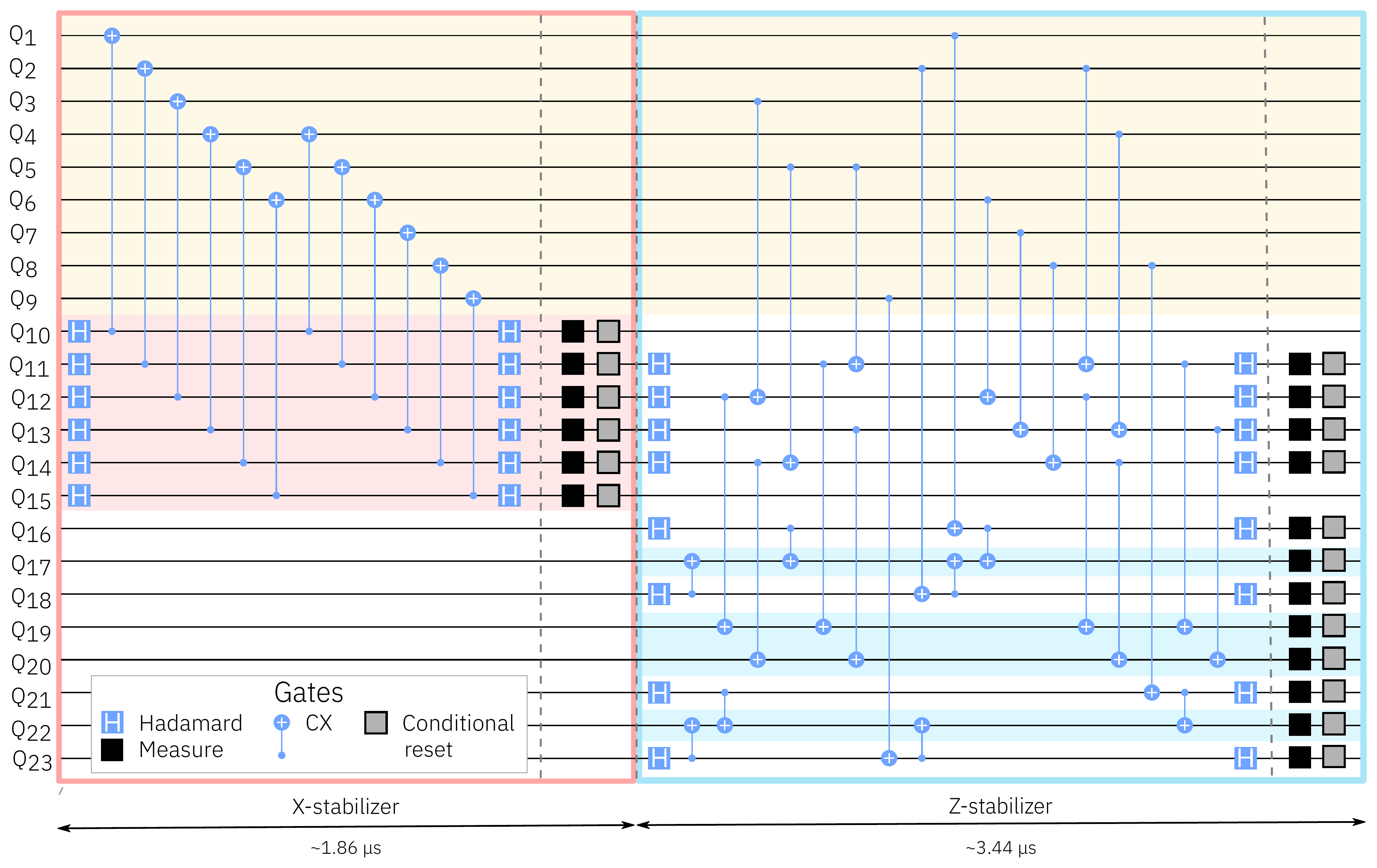}\label{fig:RoundSched_b}}}
    \caption{(a). $Z$ (blue) and $X$ (red) gauge operators (eq.~\ref{eq:gauge_Z} and ~\ref{eq:gauge_X}) mapped onto the 23 qubits required with the distance-3 heavy-hexagon code. Code qubits $(Q_1 - Q_9)$ are shown in yellow, syndrome qubits $(Q_{17},Q_{19},Q_{20},Q_{22})$ used for $Z$ stabilizers in blue, and flag qubits and syndromes used in $X$ stabilizers in white. The order and direction that CX gates are applied within each sub-section (0 to 4) are denoted by the numbered arrows. (b) Circuit diagram of one syndrome measurement round, including both $X$ and $Z$ stabilizers. The circuit diagram illustrates permitted parallelization of gate operations as set by barriers; and as each two-qubit gate duration differs, the final gate scheduling is determined with a standard circuit transpilation pass. } 
    \label{fig:RoundSched}
\end{figure*}

The heavy-hexagon code example we consider is an $n=9$ qubit code with minimum distance $d=3$ \cite{chamberlandTopologicalSubsystemCodes2020}. The $Z$ and $X$ gauge (see Fig.~\ref{fig:RoundSched_a}) and stabilizer groups are generated by
\begin{align}
{\cal G}_Z & = \langle Z_1Z_2, Z_2Z_3Z_5Z_6, Z_4Z_5Z_7Z_8, Z_8Z_9\rangle \label{eq:gauge_Z}\\
{\cal G}_X & = \langle X_1X_4, X_2X_5, X_3X_6, X_4X_7, X_5X_8, X_6X_9\rangle \label{eq:gauge_X}\\
{\cal S}_Z & = \langle Z_1Z_2Z_4Z_5Z_7Z_8, Z_2Z_3Z_5Z_6Z_8Z_9\rangle \\
{\cal S}_X & = \langle X_1X_2X_4X_5, X_3X_6, X_4X_7, X_5X_6X_8X_9\rangle
\end{align}

For this work, we focus on a particular kind of FT circuit, but the same approach can be used more generally with different codes and circuits. Two sub-circuits, shown in Fig.~\ref{fig:RoundSched}(b), are constructed to measure the $X$- and $Z$-gauge operators. The $Z$-gauge measurement circuit also acquires useful information by measuring flag qubits. 

We prepare code states in the logical $\ket{0}$ ($\ket{+}$) state by first preparing nine qubits in the $\ket{0}^{\otimes9}$ ($\ket{+}^{\otimes9}$) state and measuring the $X$-gauge ($Z$-gauge). We then perform $r$ rounds of syndrome measurement, where a round consists of a $Z$-gauge measurement followed by an $X$-gauge measurement (respectively, $X$-gauge followed by $Z$-gauge). Finally, we readout all nine code qubits in the $Z$ ($X$) basis. We perform the same experiments for initial logical states $\ket{1}$ and $\ket{-}$ as well, by simply initializing the nine qubits in $\ket{1}^{\otimes9}$ and $\ket{-}^{\otimes9}$ instead.

\section{Decoding algorithms}
\label{sec:dec_alg}

In the setting of FT quantum computing, a decoder is an algorithm that takes as input syndrome measurements from an error correcting code and outputs a correction to the qubits or measurement data. In this section we describe two decoding algorithms: perfect matching decoding and maximum likelihood decoding.

\subsection{The decoding hypergraph}
\label{sec:dec_hypergraph}

\begin{figure*}[t]
    \subfloat[\centering]{\includegraphics[width=0.45\textwidth]{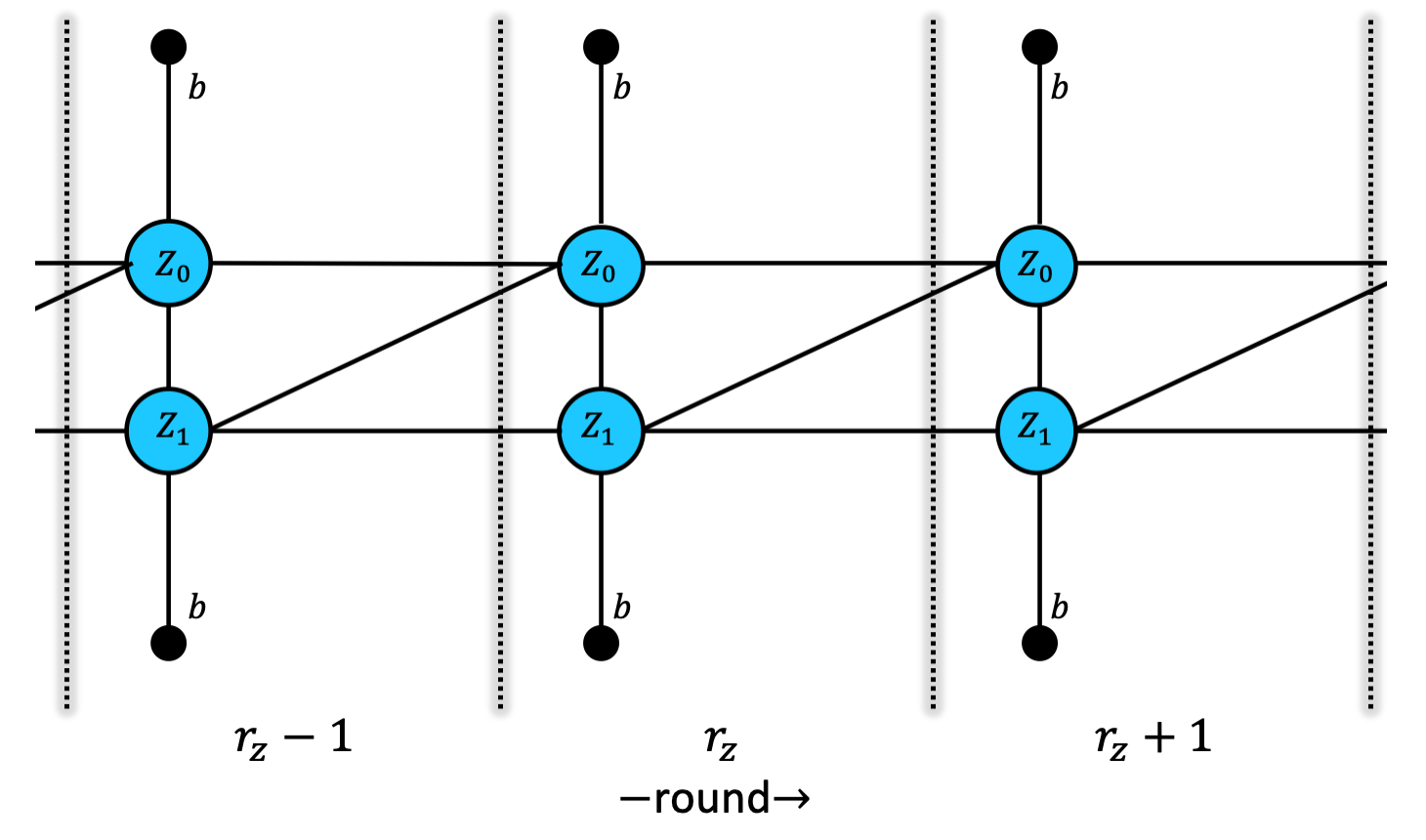}\label{fig:graph_a}} \hfil
    \subfloat[\centering]{\includegraphics[width=0.45\textwidth]{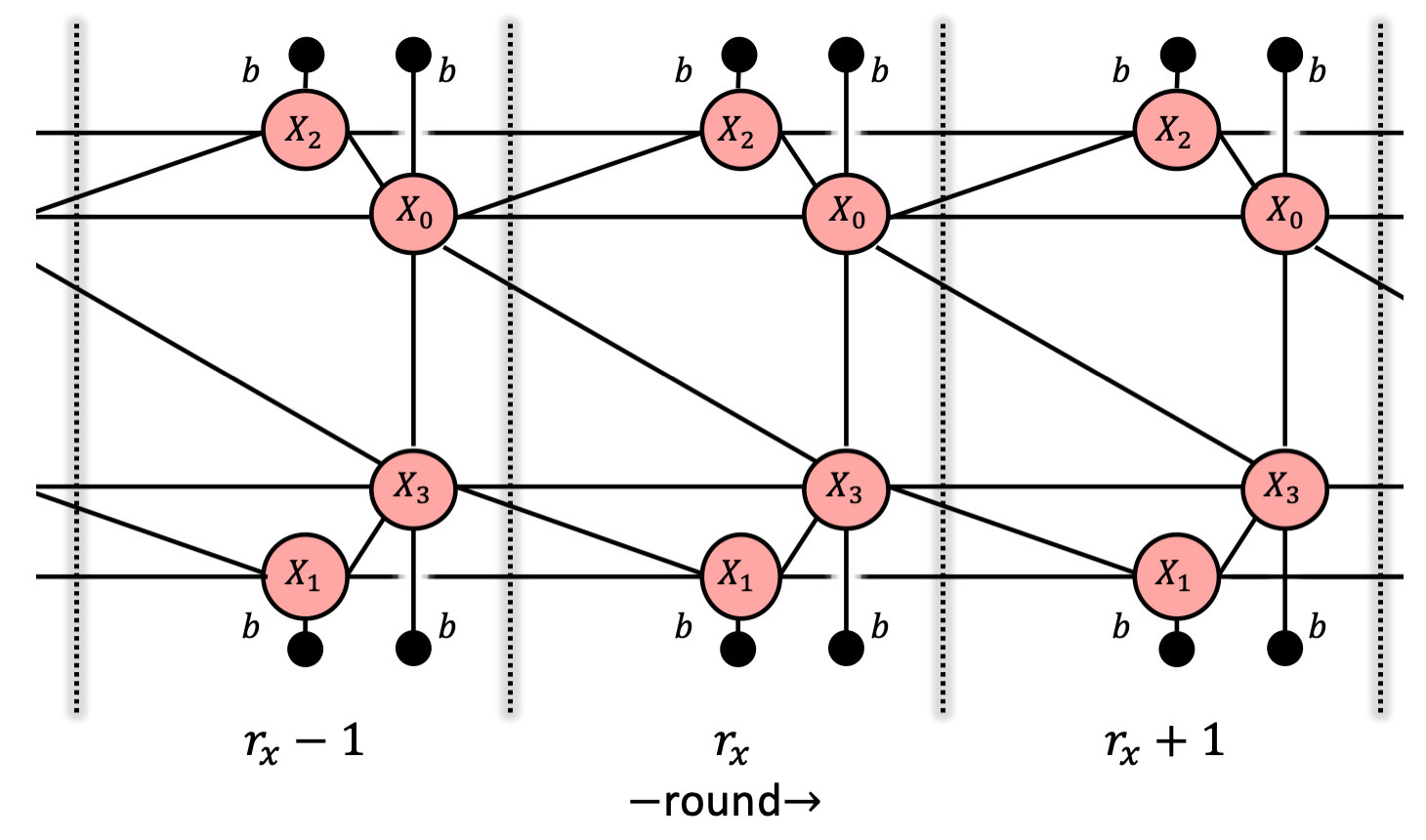}\label{fig:graph_b}}
    
    \subfloat[\centering]{\includegraphics[width=0.45\textwidth]{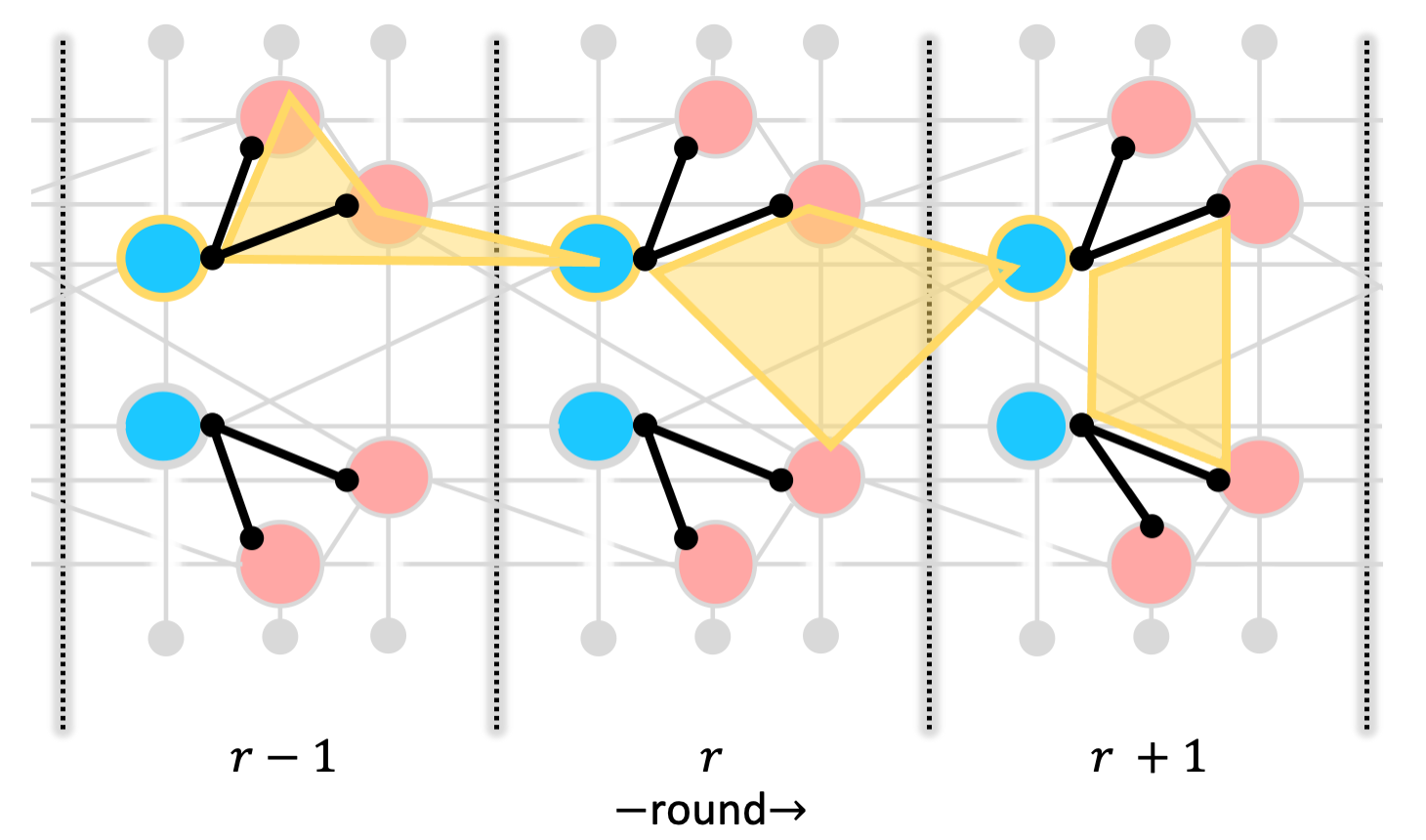}\label{fig:graph_c}} \hfil
    \subfloat[\centering]{\includegraphics[width=0.45\textwidth]{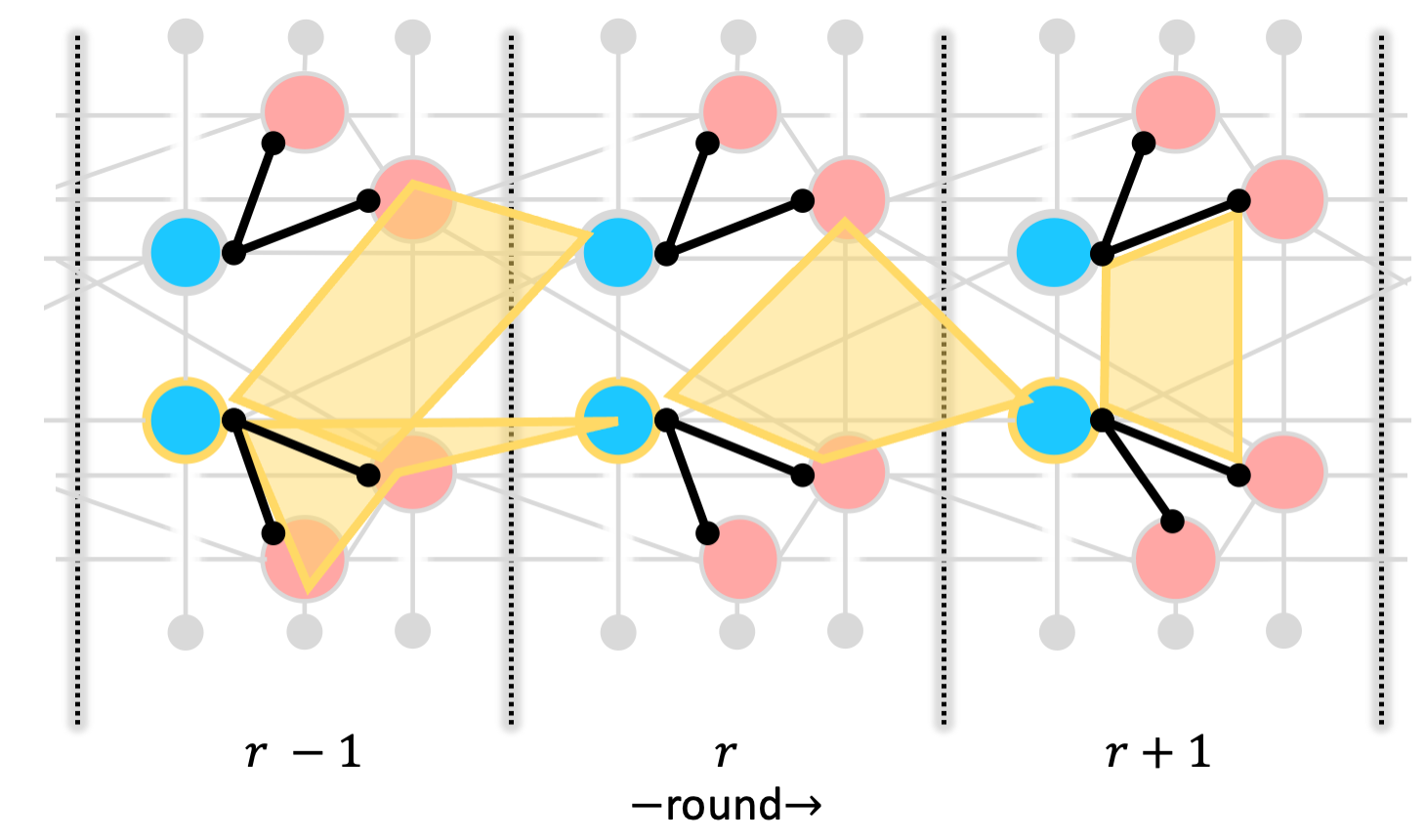}\label{fig:graph_d}}
\caption{
Decoding graphs for three rounds of (a) $Z$ and (b) $X$ stabilizer measurements for correcting $X$ and $Z$ errors, respectively, on the $d=3$ heavy-hexagon code with circuit-level noise. The blue (a) and red (b) nodes in the graph correspond to stabilizers, and the black nodes correspond to the boundary. Node labels are defined by the stabilizer measurement ($Z$ or $X$), along with a subscripts indexing the stabilizer, and superscripts denoting the round.
(c) Black edges, arising from Pauli $Y$ errors, connect the two graphs in (a) and (b). The three, size-4 hyperedges involving the top $Z$-stabilizer (gold outline).
(d) The four, size-4 hyperedges involving the bottom $Z$-stabilizer (gold outline).
}
\label{fig:graph}
\end{figure*}

The decoding hypergraph \cite{chen2021calibrated} is a concise description of the information gathered by a FT circuit and made available to a decoding algorithm. It consists of a set of vertices, or error-sensitive events, $V$, and a set of hyperedges $E$, which encode the correlations between events caused by errors in the circuit. Fig.~\ref{fig:graph} depicts parts of the decoding hypergraph for our experiment.

Constructing the decoding hypergraph for stabilizer circuits with Pauli noise can be done using standard Gottesman-Knill simulations \cite{gottesman1998heisenberg}. First, an error-sensitive event is created for each measurement that is deterministic in the error-free circuit. A deterministic measurement outcome $m\in\{0,1\}$ is determined by some function of previous measurement outcomes $\mathcal{M}$, or $m=F_m(\mathcal{M})$, where $F_m$ can be found by simulation of the error-free circuit. The value of the associated error-sensitive event is defined to be $m-F_m(\mathcal{M})\text{ (mod 2)}$, which is zero (also called trivial) in the absence of errors. Thus, observing a non-zero (also called non-trivial) error-sensitive event implies the circuit suffered at least one error. In our circuits, error-sensitive events are either flag qubit measurements or the difference of subsequent measurements of the same stabilizer. Note that measurements of a stabilizer are found by adding together measurements of the constituent gauge operators \cite{chamberlandTopologicalSubsystemCodes2020}.

Next, hyperedges are added by considering circuit faults. Our model contains a fault probability $p_C$ for each of several circuit components 
\begin{equation}
C\in\{\mathrm{cx},\mathrm{h},\mathrm{id},\mathrm{idm},\mathrm{x},\mathrm{y},\mathrm{z},\mathrm{measure},\mathrm{initialize},\mathrm{reset}\}.
\end{equation}
Here we distinguish the identity operation $\mathrm{id}$ on qubits during a time when other qubits are undergoing unitary gates, from the identity operation $\mathrm{idm}$ on qubits when others are undergoing measurement and reset. We $\mathrm{reset}$ qubits after they are measured, while we $\mathrm{initialize}$ qubits that have not been used in the experiment yet (see \hyperref[supp:experiment-details]{Supp.~\nameref{supp:experiment-details}} for more detail). Numerical values for $p_C$ are listed in \hyperref[supp:experiment-details]{Supp.~\nameref{supp:experiment-details}}. 

For initialization and reset errors, a Pauli $X$ is applied with the respective probabilities after the ideal state preparation. For measurement errors, Pauli $X$ is applied with probability $p_{\mathrm{measure}}$ before the ideal measurement. A one-qubit unitary gate (two-qubit gate) $C$ suffers with probability $p_C$ one of the three (fifteen) non-identity one-qubit (two-qubit) Pauli errors following the ideal gate. There is an equal chance of any of the three (fifteen) Pauli errors occurring.

When a single fault occurs in the circuit, it causes some subset of error-sensitive events to be non-trivial. This set of error-sensitive events becomes a hyperedge. The set of all hyperedges is $E$. Two different faults may lead to the same hyperedge, so each hyperedge may be viewed as representing a set of faults, each of which individually causes the events in the hyperedge to be non-trivial. Associated with each hyperedge is a probability, which, at first order, is the sum of the probabilities of faults in the set.

A fault may also lead to an error which, propagated to the end of the circuit, anti-commutes with one or more of the code's logical operators (say it has $k$ logical qubits and a basis of $2k$ logical operators). We can keep track of which logical operators anti-commute with the error using a vector from $\mathbb{Z}_2^{2k}$. Thus, each hyperedge $h$ is also labeled by one of these vectors $\gamma_h\in\mathbb{Z}_2^{2k}$, called a \emph{logical label}. Note that if the code has distance at least three, each hyperedge has a unique logical label \footnote{In the distance 2 case of \cite{chen2021calibrated}, the hyperedges that did not have a unique logical label were called \emph{ambiguous}, and were the motivation for a partial post-selection scheme.}.

Lastly, we note that a decoding algorithm can choose to simplify the decoding hypergraph in various ways. One way that we always employ here is the process of deflagging \footnote{Flag measurements from qubits 16, 18, 21, 23 are simply ignored with no corrections applied. If flag 11 is non-trivial and 12 trivial, apply $Z$ to 2. If 12 is non-trivial and 11 trivial, apply $Z$ to qubit 6. If flag 13 is non-trivial and 14 trivial, apply $Z$ to qubit 7. If 14 is non-trivial and 13 trivial, apply $Z$ to qubit 8. See \cite{chen2021calibrated} for details on why this is sufficient for fault-tolerance.}. This means that instead of including error-sensitive events from the flag qubit measurements directly, we use the flag information to immediately (before any more gates are applied) apply virtual Pauli $Z$ corrections and adjust subsequent error-sensitive events accordingly. Hyperedges for the deflagged hypergraph can be found through stabilizer simulation incorporating the $Z$ corrections.

\subsection{Perfect Matching Decoding}
\label{sec:matching}
Considering $X$ and $Z$ errors separately, the problem of finding a minimum weight error correction for the surface code can be reduced to finding a minimum weight perfect matching in a graph \cite{Dennis2002TopologicalQM}. Matching decoders continue to be studied because of their practicality \cite{Fowler2012matching} and broad applicability \cite{higgottPyMatchingPythonPackage2021, Dua2022fractal}. In this section, we describe the matching decoder for our distance-3 heavy-hexagon code.

The decoding graphs, one for the $X$-errors (Fig.~\ref{fig:graph_a}) and one for the $Z$-errors (Fig.~\ref{fig:graph_b}), for minimum weight perfect matching are in fact subgraphs of the decoding hypergraph in the previous section. Let us focus here on the graph for correcting $X$-errors, since the $Z$-error graph is analogous. In this case, from the decoding hypergraph we keep nodes $V_Z$ corresponding to (the difference of subsequent) $Z$-stabilizer measurements and edges (i.e.~hyperedges with size two) between them. Additionally, a boundary vertex $b$ is created, and size-one hyperedges of the form $\{v\}$ with $v\in V_Z$, are represented by including edges $\{v,b\}$. All edges in the $X$-error graph inherit probabilities and logical labels from their corresponding hyperedges (see Table \ref{tab:edgeweZ} (\ref{tab:edgeweX}) for $X$ ($Z$)-error edge data for 2-round experiment).

A perfect matching algorithm takes a graph with weighted edges and an even-sized set of highlighted nodes, and returns a set of edges in the graph that connects all highlighted nodes in pairs and has minimum total weight among all such edge sets. In our case, highlighted nodes are the non-trivial error-sensitive events (if there are an odd number, the boundary node is also highlighted), and edge weights are either chosen to all be one (uniform method) or set as $w_e=\log\left((1-p_e)/p_e\right)$, where $p_e$ is the edge probability (analytic method). The latter choice means that the total weight of an edge set is equal to the log-likelihood of that set, and minimum weight perfect matching tries to maximize this likelihood over the edges in the graph.

Given a minimum weight perfect matching, one can use the logical labels of the edges in the matching to decide on a correction to the logical state. Alternatively, the $X$-error ($Z$-error) graph for the matching decoder is such that each edge can be associated to a code qubit (or a meausurement error), such that including an edge in the matching implies an $X$ ($Z$) correction should be applied to the corresponding qubit.

\subsection{Maximum Likelihood Decoding}

Maximum likelihood decoding (MLD) is an optimal, albeit non-scalable, method for decoding quantum error-correcting codes. In its original conception, MLD was applied to phenomenological noise models where errors occur only just before syndromes are measured \cite{bravyiEfficientAlgorithmsMaximum2014,bravyiDoubledColorCodes2015}. This of course ignores the more realistic case where errors can propagate through the syndrome measurement circuitry. More recently, MLD has been extended to include circuit noise \cite{heimOptimalCircuitLevelDecoding2016,pryadko2020maximum}. Here, we describe how MLD corrects circuit noise using the decoding hypergraph.

MLD deduces the most likely logical correction given an observation of the error-sensitive events. This is done by calculating the probability distribution $\text{Pr}[\beta\gamma]$, where $\beta\in\mathbb{Z}_2^{|V|}$ represents error-sensitive events and $\gamma\in\mathbb{Z}_2^{2k}$ represents a logical correction.

We can calculate $\text{Pr}[\beta\gamma]$ by including each hyperedge from the decoding hypergraph, starting from the zero-error distribution, i.e.~ $\text{Pr}[0^{|V|}0^{2k}]=1$. If hyperedge $h$ has some probability $p_h$ of occurring, independent of any other hyperedge, we include $h$ by performing the update
\begin{equation}\label{eq:ml_update}
\text{Pr}[\beta\gamma]\leftarrow(1-p_h)\text{Pr}[\beta\gamma]+p_h\text{Pr}[(\beta\oplus\beta_h)(\gamma\oplus\gamma_h)],
\end{equation}
where $\beta_h\in\mathbb{Z}_2^{|V|}$ is just a binary vector representation of the hyperedge. This update should be applied once for every hyperedge in $E$.

Once $\text{Pr}[\beta\gamma]$ is calculated, we can use it to deduce the best logical correction. If $\beta^*\in\mathbb{Z}_2^{|V|}$ is observed in a run of the experiment,
\begin{equation}\label{eq:ml_correction}
\gamma^*=\text{argmax}_\gamma\text{Pr}[\beta^*\gamma]
\end{equation}
indicates how measurements of the logical operators should be corrected. For more details on specific implementations of MLD, refer to \hyperref[supp:MLD]{Supp.~\nameref{supp:MLD}}.

\section{Experimental Discussion}
\begin{figure}
    \centering
    \subfloat[\centering]{\includegraphics[width=0.45\textwidth]{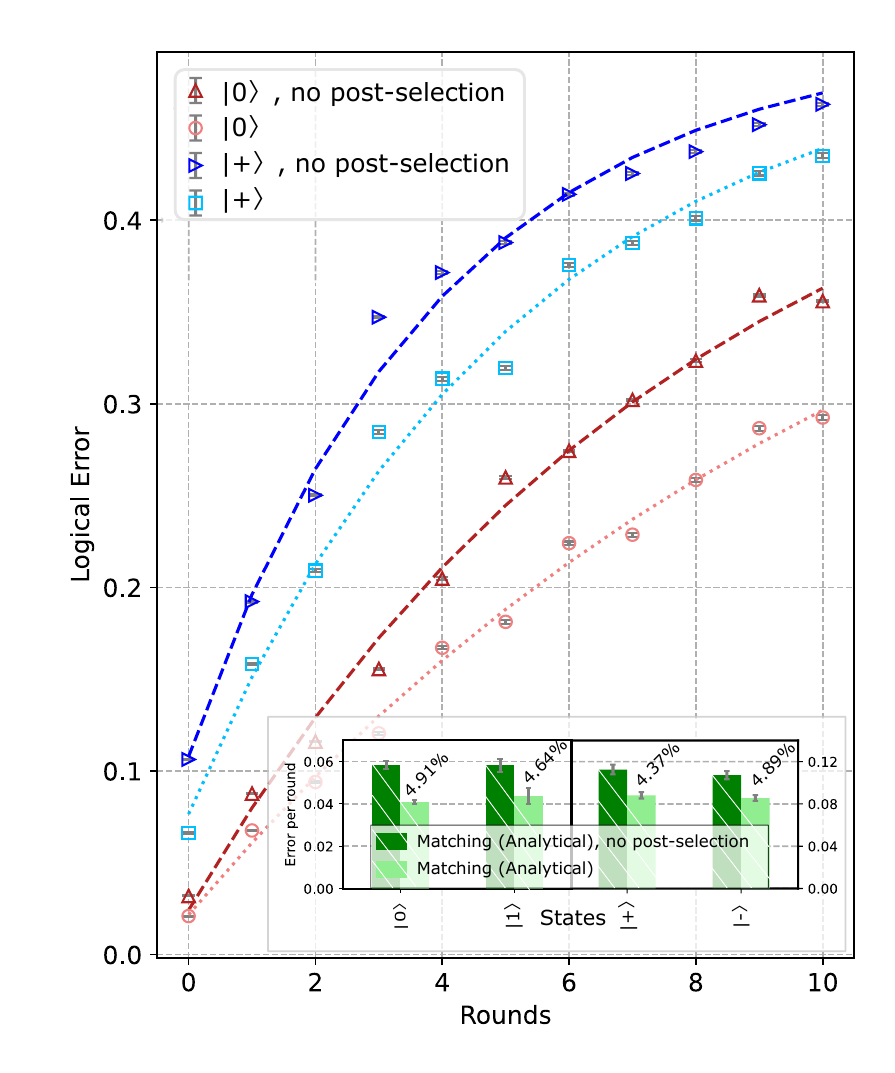}\label{fig:RoundvsError_a}} \\
    \subfloat[\centering]{\includegraphics[width=0.48\textwidth]{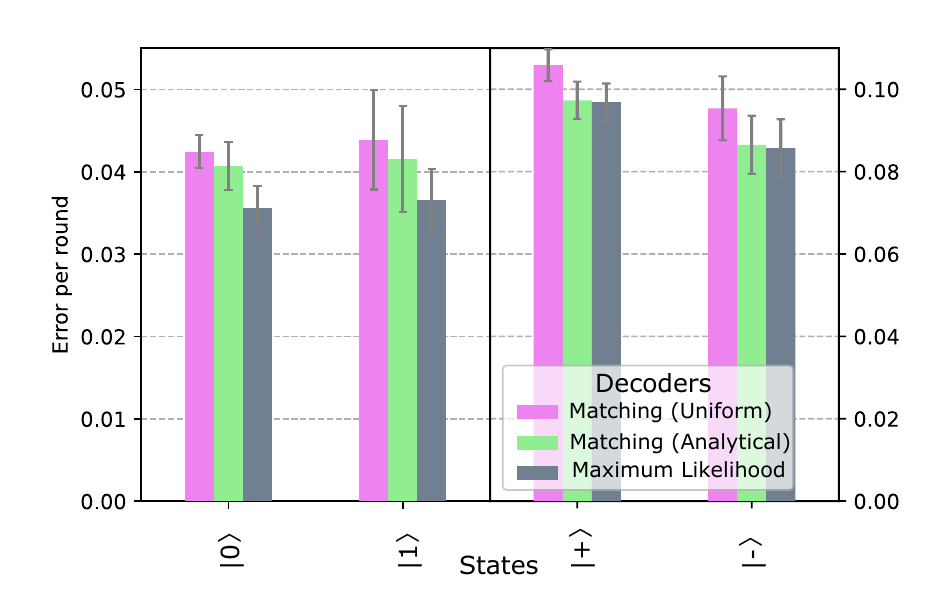}\label{fig:RoundvsError_b}}
    \caption{(a) Logical error, from matching analytical decoding, vs. number of syndrome measurement rounds $r$, where one round includes both $Z$- and $X$-basis measurements. Error bars denote sampling error of each run (500,000 shots). Dashed line fits of error (including all rounds) yield error per round plotted in (a-inset). Applying the same decoding method on leakage-post-selected data (rejection rate per round quoted above post-selected error rates in inset), shows substantial reduction in overall error. See \hyperref[supp:post-select-method]{Supp.~\nameref{supp:post-select-method}} for details. (b) Comparison of fitted error per round (including up to 4 rounds in the fit) for all four logical states using matching uniform, matching analytical, and maximum likelihood decoders. Error bars here represent one standard deviation on the fitted rate.} 
    \label{fig:RoundvsError}
\end{figure}

For this demonstration we use $ibm\_peekskill$, a 27 qubit IBM Quantum Falcon processor \footnote{$ibm\_peekskill$ v2.0.0, an IBM Quantum Falcon R8 processor.  https://quantum-computing.ibm.com/, accessed  Jan. 2022} whose coupling map enables a distance-3 heavy-hexagon code, see Fig.~\ref{fig:RoundSched}.  
The total time for qubit measurement and subsequent real-time conditional reset, for each round, takes 768ns and is the same for all qubits. All syndrome measurements and resets occur simultaneously for improved performance. As the final step, a simple $X_\pi$-$X_\pi$ dynamical decoupling sequence is added to all code qubits during their respective idling periods. 

Qubit leakage is a significant reason why the Pauli depolarizing error-model assumed by the decoder design might be inaccurate. It is sometimes possible to detect whether a qubit has leaked when it is measured. Thus, we can post-select on runs of the experiment when leakage has not been detected, similar to \cite{Krinner2021RealizingRQ}. See \hyperref[supp:post-select-method]{Supp.~\nameref{supp:post-select-method}} for more information on the post-selection method.

In Fig.~\ref{fig:RoundvsError_a}, we initialize the logical state $\vert0\rangle$ ($\vert+\rangle$), and apply $r$ syndrome measurement rounds, where one round includes both $X$ and $Z$ stabilizers (total time of approximately 5.3$\mu$s per round, Fig.~\ref{fig:RoundSched_b}). Using analytical perfect matching decoding on the full data set (500,000 shots per run), we extract the logical errors in Fig.~\ref{fig:RoundvsError_a}, red (blue) triangles. Details of optimized parameters used in analytical perfect matching decoding can be found in \hyperref[supp:experiment-details]{Supp.~\nameref{supp:experiment-details}}.  Fitting the full decay curves (eq.~\ref{eq:pfail_fit}) up to 10 rounds, we extract logical error per round without post-selection in Fig.~\ref{fig:RoundvsError_a}-inset of 0.059(2) (0.058(3)) for $\vert0\rangle$ ($\vert1\rangle$) and 0.113(5) (0.107(4)) for $\vert+\rangle$ ($\vert-\rangle$).

Applying the same decoding method on leakage-post-selected data reduces logical errors in Fig.~\ref{fig:RoundvsError_a}, and leads to fitted error rates of 0.041(1) (0.044(4)) for $\vert0\rangle$ ($\vert1\rangle$) and 0.088(3) (0.085(3)) for $\vert+\rangle$ ($\vert-\rangle$).  Rejection rates per round from post-selection are marked in the inset above corresponding error rate fits. See \hyperref[supp:post-select-method]{Supp.~\nameref{supp:post-select-method}} for details.

In Fig.~\ref{fig:RoundvsError_b}, we compare the logical error per round from the post-selected data sets using the three decoders described previously in Section~\ref{sec:dec_alg}. To match the current limitation of MLD to $r=4$ rounds, all error rates in Fig.~\ref{fig:RoundvsError_b} were fit using only rounds $r=0$ to 4, with error bars here denoting one standard deviation of the fitting parameter. We observe a consistent improvement in decoding moving from matching uniform (pink), to matching analytical (green), to maximum likelihood (grey). A quantitative comparison between the three decoders for all four logical states at $r=2$ rounds is provided in \hyperref[supp:neq2]{Supp.~\nameref{supp:neq2}}.

\section{Conclusions and outlook}
The results presented in this work highlight the importance of the joint progress of quantum hardware, both in size and quality, and classical information processing, both concurrent with circuit execution and asynchronous to it, as described with the studied decoders. Our experiments incorporate mid-circuit measurements and conditional operations as part of a quantum error correction protocol. These technical capabilities will serve as foundational elements for further enhancement of the role of dynamic circuits in quantum error correction, for example towards real-time correction and other feed-forward operations that could be critical for large-scale FT computations. We also show how experimental platforms for quantum error correction of this size and capabilities can trigger new ideas towards more robust decoders. Our comparison between a perfect matching and a maximum likelihood decoder sets a promising starting point towards the understanding of the trade-off between decoder scalability versus performance in the presence of experimental noise.

All these key components will play a crucial role in larger distance codes, where the quality of the real-time operations (qubit conditional reset and leakage removal, teleportation protocols for logical gates, and decoding), along with device noise levels, will determine the performance of the code, potentially enabling the demonstration of logical error suppression with increased code distance.

\section*{Acknowledgements}
The device was designed and fabricated internally at IBM. We acknowledge the use of IBM Quantum services for this work, and these results were enabled by the work of the IBM Quantum software and hardware teams. This code demonstration work was supported by IARPA under LogiQ (contract W911NF-16-1-0114). All statements of fact, opinion or conclusions contained herein are
those of the authors and should not be construed as representing the official views or
policies of the US Government.

\clearpage
\newpage

\onecolumngrid
\beginsupplement
\section*{Supplement}
\renewcommand{\theequation}{S.\arabic{equation}}
\setcounter{equation}{0}

\subsection{Minimum weight perfect matching edge probabilities}

Here we list the edge probabilities for the decoder graphs used in minimum weight perfect matching. Note that for experiments on logical $\ket{0}$ and $\ket{1}$, we need only correct $X$ errors and so just use the $Z$ stabilizers, Fig.~\ref{fig:graph_a}. For experiments on logical $\ket{+}$ and $\ket{-}$, we need only correct $Z$ errors with the graph in Fig.~\ref{fig:graph_b}. Edge weighs are given for the logical $\ket{0}$ and $\ket{+}$ 2-round experiments in the respective Tables~\ref{tab:edgeweZ} and \ref{tab:edgeweX}.

\begin{table}[b]

\begin{tabular}{c|c|c}
\toprule
Edge $e$ & Qubit $Q(e)$ & First-order edge flip probability $\tilde{p}_e$ \\
\hline
(${z}_{0}^{1}$, ${z}_{1}^{1}$) & {2} & $44/15 p_{cx}+14/3 p_{id}+3 p_{init}+2 p_{idm}$ \\
(${z}_{0}^{1}$, $b$) & {1} & $44/15 p_{cx}+6 p_{id}+3 p_{init}+2 p_{idm}$ \\
(${z}_{0}^{1}$, ${z}_{0}^{2}$) & {$\emptyset$} & $88/15 p_{cx}+4/3 p_{id}+2 p_{init}+2 p_{measure}$ \\
(${z}_{1}^{1}$, $b$) & {3} & $44/15 p_{cx}+4 p_{id}+3 p_{init}+2 p_{idm}$ \\
(${z}_{1}^{1}$, ${z}_{0}^{2}$) & {2} & $8/5 p_{cx}$ \\
(${z}_{1}^{1}$, ${z}_{1}^{2}$) & {$\emptyset$} & $88/15 p_{cx}+4/3 p_{id}+2 p_{init}+2 p_{measure}$ \\
(${z}_{0}^{2}$, ${z}_{1}^{2}$) & {2} & $56/15 p_{cx}+22/3 p_{id}+4 p_{idm}$ \\
(${z}_{0}^{2}$, $b$) & {1} & $56/15 p_{cx}+28/3 p_{id}+4 p_{idm}$ \\
(${z}_{0}^{2}$, ${z}_{0}^{3}$) & {$\emptyset$} & $88/15 p_{cx}+4/3 p_{id}+2 p_{measure}+2 p_{reset}$ \\
(${z}_{1}^{2}$, $b$) & {3} & $56/15 p_{cx}+28/3 p_{id}+4 p_{idm}$ \\
(${z}_{1}^{2}$, ${z}_{0}^{3}$) & {2} & $8/5 p_{cx}$ \\
(${z}_{1}^{2}$, ${z}_{1}^{3}$) & {$\emptyset$} & $88/15 p_{cx}+4/3 p_{id}+2 p_{measure}+2 p_{reset}$ \\
(${z}_{0}^{3}$, ${z}_{1}^{3}$) & {2} & $44/15 p_{cx}+4 p_{id}+2 p_{idm}+3 p_{measure}$ \\
(${z}_{0}^{3}$, $b$) & {1} & $44/15 p_{cx}+14/3 p_{id}+2 p_{idm}+3 p_{measure}$ \\
(${z}_{1}^{3}$, $b$) & {3} & $44/15 p_{cx}+20/3 p_{id}+2 p_{idm}+3 p_{measure}$ \\
\bottomrule
\end{tabular}

\caption{Edge data for the $X$-error decoding graph shown in Figure~\ref{fig:graph_a}. Here $z_s^t$ indicates the $s^{\text{th}}$ $Z$-stabilizer at time $t$, as in Fig.~\ref{fig:graph_a}, and $b$ is the boundary node. If an edge $e$ is chosen by the matching algorithm, a Pauli $X$ correction is applied to qubit $Q(e)$ if it is not $\emptyset$.}
\label{tab:edgeweZ}
\end{table}

\begin{table}
\centering

\begin{tabular}{c|c|c}
\toprule
Edge $e$ & Qubits $Q(e)$ & First-order edge flip probability $\tilde{p}_e$ \\
\hline
(${x}_{0}^{1}$, ${x}_{2}^{1}$) & {4} & $4/3 p_{H}+8/5 p_{cx}+8/3 p_{id}+ p_{init}+2/3 p_{idm}+ p_{measure}$ \\
(${x}_{0}^{1}$, ${x}_{3}^{1}$) & {5} & $2 p_{H}+12/5 p_{cx}+2 p_{id}+3 p_{init}+2/3 p_{idm}$ \\
(${x}_{0}^{1}$, $b$) & {1} & $10/3 p_{H}+4 p_{cx}+16/3 p_{id}+4 p_{init}+4/3 p_{idm}+ p_{measure}$ \\
(${x}_{0}^{1}$, ${x}_{0}^{2}$) & {$\emptyset$} & $8/3 p_{H}+16/15 p_{cx}+ p_{init}+2 p_{measure}+ p_{reset}$ \\
(${x}_{0}^{1}$, ${x}_{2}^{2}$) & {4} & $8/15 p_{cx}$ \\
(${x}_{0}^{1}$, ${x}_{3}^{2}$) & {5} & $8/15 p_{cx}$ \\
(${x}_{1}^{1}$, ${x}_{3}^{1}$) & {6} & $4/3 p_{H}+16/15 p_{cx}+10/3 p_{id}+ p_{init}+2/3 p_{idm}+ p_{measure}$ \\
(${x}_{1}^{1}$, $b$) & {3} & $4/3 p_{H}+8/5 p_{cx}+8/3 p_{id}+2 p_{init}+2/3 p_{idm}$ \\
(${x}_{1}^{1}$, ${x}_{1}^{2}$) & {$\emptyset$} & $4/3 p_{H}+8/15 p_{cx}+ p_{measure}+ p_{reset}$ \\
(${x}_{1}^{1}$, ${x}_{3}^{2}$) & {6} & $8/15 p_{cx}$ \\
(${x}_{2}^{1}$, $b$) & {7} & $4/3 p_{H}+16/15 p_{cx}+10/3 p_{id}+2 p_{init}+2/3 p_{idm}$ \\
(${x}_{2}^{1}$, ${x}_{2}^{2}$) & {$\emptyset$} & $4/3 p_{H}+8/15 p_{cx}+ p_{measure}+ p_{reset}$ \\
(${x}_{3}^{1}$, $b$) & {8} & $10/3 p_{H}+52/15 p_{cx}+22/3 p_{id}+4 p_{init}+4/3 p_{idm}+ p_{measure}$ \\
(${x}_{3}^{1}$, ${x}_{3}^{2}$) & {$\emptyset$} & $8/3 p_{H}+16/15 p_{cx}+ p_{init}+2 p_{measure}+ p_{reset}$ \\
(${x}_{0}^{2}$, ${x}_{2}^{2}$) & {4} & $2/3 p_{H}+28/15 p_{cx}+8/3 p_{id}+4/3 p_{idm}+ p_{measure}$ \\
(${x}_{0}^{2}$, ${x}_{3}^{2}$) & {5} & $4/3 p_{H}+8/3 p_{cx}+2 p_{id}+4/3 p_{idm}+2 p_{reset}$ \\
(${x}_{0}^{2}$, $b$) & {1} & $2 p_{H}+68/15 p_{cx}+20/3 p_{id}+8/3 p_{idm}+ p_{measure}+2 p_{reset}$ \\
(${x}_{0}^{2}$, ${x}_{0}^{3}$) & {$\emptyset$} & $8/3 p_{H}+16/15 p_{cx}+2 p_{measure}+2 p_{reset}$ \\
(${x}_{0}^{2}$, ${x}_{2}^{3}$) & {4} & $8/15 p_{cx}$ \\
(${x}_{0}^{2}$, ${x}_{3}^{3}$) & {5} & $8/15 p_{cx}$ \\
(${x}_{1}^{2}$, ${x}_{3}^{2}$) & {6} & $2/3 p_{H}+4/3 p_{cx}+10/3 p_{id}+4/3 p_{idm}+ p_{measure}$ \\
(${x}_{1}^{2}$, $b$) & {3} & $2/3 p_{H}+28/15 p_{cx}+10/3 p_{id}+4/3 p_{idm}+ p_{reset}$ \\
(${x}_{1}^{2}$, ${x}_{1}^{3}$) & {$\emptyset$} & $4/3 p_{H}+8/15 p_{cx}+ p_{measure}+ p_{reset}$ \\
(${x}_{1}^{2}$, ${x}_{3}^{3}$) & {6} & $8/15 p_{cx}$ \\
(${x}_{2}^{2}$, $b$) & {7} & $2/3 p_{H}+4/3 p_{cx}+10/3 p_{id}+4/3 p_{idm}+ p_{reset}$ \\
(${x}_{2}^{2}$, ${x}_{2}^{3}$) & {$\emptyset$} & $4/3 p_{H}+8/15 p_{cx}+ p_{measure}+ p_{reset}$ \\
(${x}_{3}^{2}$, $b$) & {8} & $2 p_{H}+4 p_{cx}+22/3 p_{id}+8/3 p_{idm}+ p_{measure}+2 p_{reset}$ \\
(${x}_{3}^{2}$, ${x}_{3}^{3}$) & {$\emptyset$} & $8/3 p_{H}+16/15 p_{cx}+2 p_{measure}+2 p_{reset}$ \\
(${x}_{0}^{3}$, ${x}_{2}^{3}$) & {4} & $4/3 p_{H}+8/5 p_{cx}+8/3 p_{id}+2/3 p_{idm}+2 p_{measure}$ \\
(${x}_{0}^{3}$, ${x}_{3}^{3}$) & {5} & $2 p_{H}+12/5 p_{cx}+2 p_{id}+2/3 p_{idm}+ p_{measure}+2 p_{reset}$ \\
(${x}_{0}^{3}$, $b$) & {1} & $10/3 p_{H}+4 p_{cx}+20/3 p_{id}+4/3 p_{idm}+3 p_{measure}+2 p_{reset}$ \\
(${x}_{1}^{3}$, ${x}_{3}^{3}$) & {6} & $4/3 p_{H}+16/15 p_{cx}+10/3 p_{id}+2/3 p_{idm}+2 p_{measure}$ \\
(${x}_{1}^{3}$, $b$) & {3} & $4/3 p_{H}+8/5 p_{cx}+10/3 p_{id}+2/3 p_{idm}+ p_{measure}+ p_{reset}$ \\
(${x}_{2}^{3}$, $b$) & {7} & $4/3 p_{H}+16/15 p_{cx}+8/3 p_{id}+2/3 p_{idm}+ p_{measure}+ p_{reset}$ \\
(${x}_{3}^{3}$, $b$) & {8} & $10/3 p_{H}+52/15 p_{cx}+6 p_{id}+4/3 p_{idm}+3 p_{measure}+2 p_{reset}$ \\
\bottomrule
\end{tabular}

\caption{Edge data for the $Z$-error decoding graph shown in Figure~\ref{fig:graph_a}. Here $x_s^t$ indicates the $s^{\text{th}}$ $X$-stabilizer at time $t$, as in Fig.~\ref{fig:graph_b}, and $b$ is the boundary node. If an edge $e$ is chosen by the matching algorithm, a Pauli $Z$ correction is applied to qubit $Q(e)$ if it is not $\emptyset$.}
\label{tab:edgeweX}
\end{table}

\subsection{Maximum likelihood implementations}\labelname{B}\label{supp:MLD}
There are at least two different ways to implement maximum likelihood decoding (MLD), which we call the offline and online usages of the decoder. They can differ significantly in time complexity depending on the specific application.

In the offline case, one calculates and stores the entire distribution $\text{Pr}[\beta\gamma]$ and queries it to determine the correction for each run of the circuit. The calculation takes $O(|E|2^{|V|+2k})$ time, since we must perform updates from Eq.~\eqref{eq:ml_update} to the distribution for each hyperedge in $E$. Determining a correction using Eq.~\eqref{eq:ml_correction} takes $O(2^{2k})$ time per run.

Alternatively, one can forgo storing the whole distribution, and instead calculate sparse distributions specific to each observation string $\beta^*$ in a data set. Online MLD achieves this by pruning the distribution as updates are performed, keeping only entries consistent with $\beta^*$. We imagine receiving one bit of $\beta^*$ at a time. For the $j^{\text{th}}$ bit, updates are made using Eq.~\eqref{eq:ml_update} for all hyperedges that contain bit $j$ and have not already been included. In fact, all updates for a given bit can be combined into a pre-calculated transition matrix. Since no further updates will be made to bit $j$, we can now truncate the distribution by keeping only entries $\text{Pr}[\beta\gamma]$ where $\beta_j=\beta^*_j$. 

During the course of online MLD, there is some maximum instantaneous size of the probability distribution, say $S_{\max}$, and the total time to determine a correction is $O(|V|S_{\max})$ per run. Note that $S_{\max}$ depends on the decoding hypergraph and also the order in which error-sensitive events are incorporated. It can be argued that for $\llbracket n,k\rrbracket$ codes, repeated rounds of syndrome measurements, and events incorporated chronologically, $2^{n+k}\le S_{\max}\le2^{2n}$, because hyperedges do not span more than two rounds of error-sensitive events. The online decoder is also amenable to dynamic programming, storing partially calculated probability distributions up to some moderately-sized $j$. For instance, in our analysis of three-round experiments, we store distributions up to $j=15$, while for four rounds we keep up to $j=21$. 

Since online MLD takes exponential (in $n$, the number of physical qubits in the code) time per run, if $|V|$ is small enough, the offline MLD is preferable. If $|V|$ is large but $n$ and $k$ are small (perhaps a small code experiment performing many rounds of syndrome measurements), the online decoder becomes the only feasible option. In the experiments here, online MLD becomes preferable over offline MLD for three rounds and greater.

\subsection{Simulation Details}
\labelname{C}\label{supp:simulation-details}
We obtain theoretical simulation results using stabilizer simulations of the Qiskit software stack \cite{Qiskit}. In order to faithfully estimate the performance of quantum error correction circuits on IBM Quantum Falcon systems, we performed simulations of the quantum circuits with qubits mapped onto the Falcon devices using customized error models that captures the realistic noise behavior of experimental hardware.

Circuit errors in our simulation are modeled as depolarizing errors, so that the effect for different error sources of varying strength can be captured. Noise models were built following error locations and error channels described in Section \ref{sec:dec_hypergraph}: depolarizing error model for each single and two qubit operation in the quantum circuit with error rates obtained from simultaneous randomized benchmarking (RB), measurement, initialize, and reset error in the form of bit-flip error for each of those operation, and idling error in the form of depolarizing noise. 

Using the above described error model, we define a \textit{realistic} depolarizing error model where simulations are carried out with noise parameters directly exported from $ibm\_peekskill$ (Table ~\ref{table:1Q_Benchmark} and ~\ref{table:2Q_Benchmark}), including 
\begin{itemize}
    \item specific error rates for each single and two-qubit quantum operation with depolarizing quantum channel parameter obtained from simultaneous RB according to the relation $$\epsilon_{\textnormal{gate}} = \frac{2^n - 1}{2^n} (1-\alpha_{\textnormal{gate}}),$$ where $\epsilon_{\textnormal{gate}}, n, \alpha_{\textnormal{gate}}$ represents error per gate, number of qubits in gate, and depolarizing quantum channel parameter,
    \item initialization, measurement, and reset error obtained as described in Table ~\ref{table:1Q_Benchmark},
    \item idling errors with noise strength proportional to coherence limit of the gate, where coherence limit is computed using $T_1$, $T_2$ and idle time of each qubit during the execution of each quantum operation in the circuit.
\end{itemize}

Furthermore, to demonstrate average performance of the circuit in a relatively uniform depolarizing error model, we define an \textit{average} depolarizing error model where instead of the specific error rates for different gates and qubits stated above we use average error rates throughout the entire device to define the depolarizing error channels.

Using analytical perfect matching decoder parameters $p_C =$ [0.0126, 0.000266, 0.0, 0.001, 0.002, 0.000266, 0.000266, 0.0, 0.00713, 0.0142, 0.0290] following the error locations  $C=\{\mathrm{cx},\mathrm{h},\mathrm{s},\mathrm{id},\mathrm{idm},\mathrm{x},\mathrm{y},\mathrm{z},\mathrm{measure},\mathrm{initialize}, \mathrm{reset} \}$ define in Section \ref{sec:matching}, we obtained simulated per round logical error rates for circuits with up to 10 syndrome measurement rounds as $0.059 \, (0.038)$ for logical state $|0\rangle$ and $0.152 \, (0.106)$ for logical state $|+ \rangle$ under the influence of $\textit{realistic}$ ($\textit{average}$) depolarizing error model, respectively.

\subsection{$ibm\_peekskill$ and experimental details}
\labelname{D}\label{supp:experiment-details}

\begin{figure}[h]
	\centering
	\subfloat[\centering]{{\includegraphics[height=0.3\textheight]{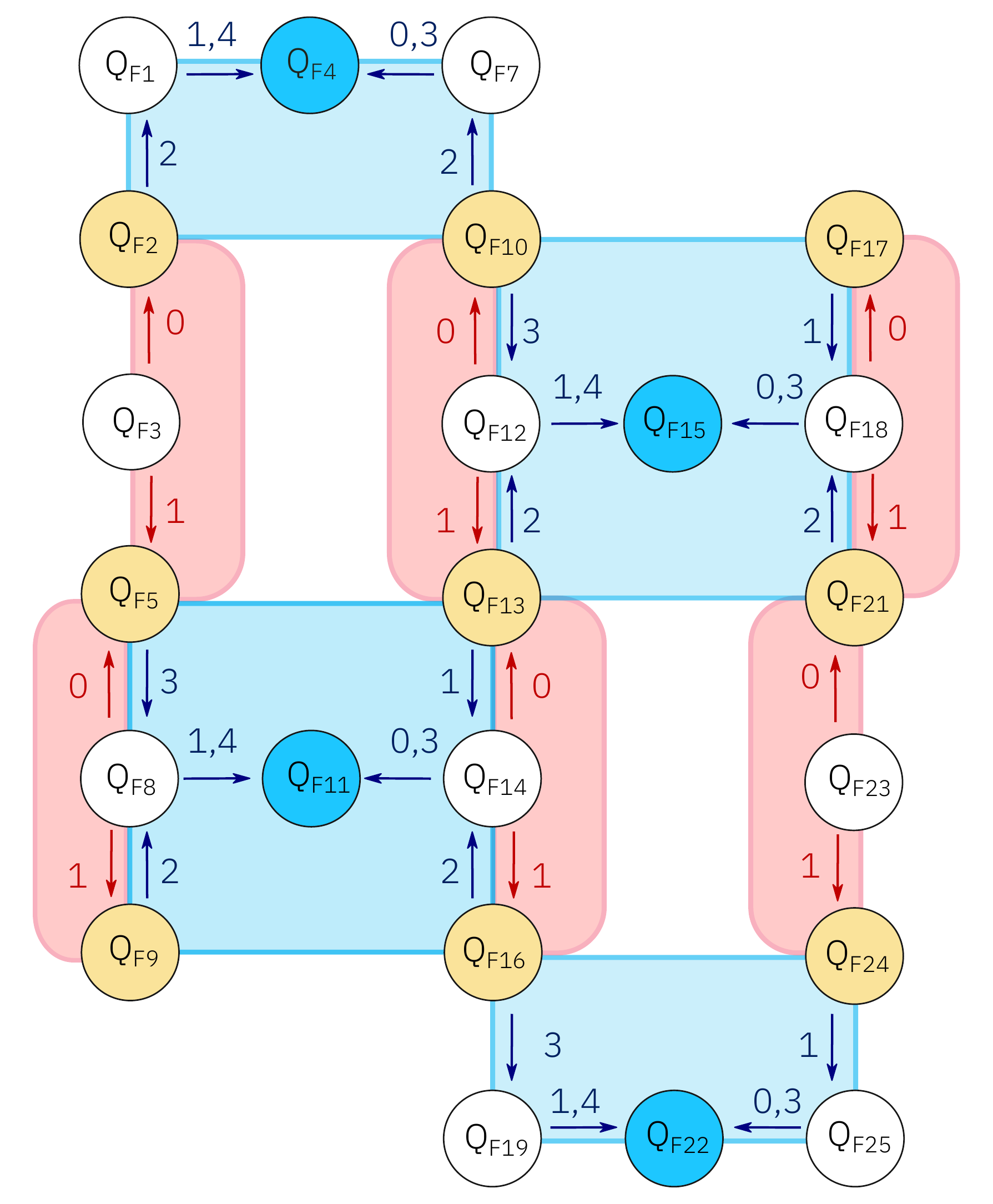}\label{fig:Fnum_ZZa}}}
	\qquad
	\subfloat[\centering]{{\includegraphics[height=0.3\textheight]{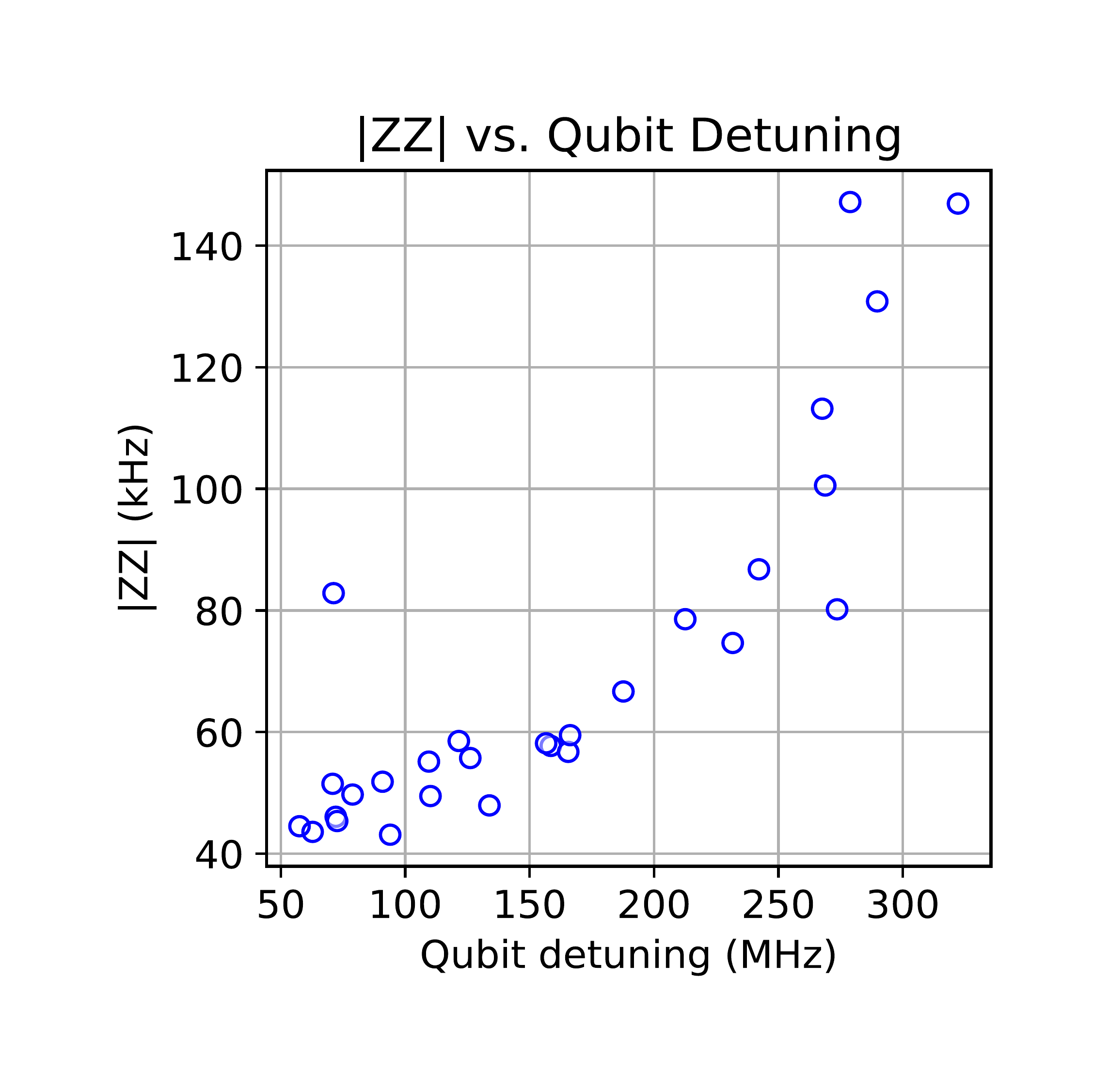}\label{fig:Fnum_ZZb}}}
			\caption{(a) Translation of Fig.~\ref{fig:RoundSched_a} qubit numbering ($Q_N$) to standard IBM-Falcon numbering($Q_{FN}$). (b) Static ZZ between all connected qubits pairs versus detuning between qubits. Median qubit anharmonicity, see Table ~\ref{table:1Q_Benchmark} for breakdown, is  -345 MHz}
	\label{fig:Fnum_ZZ}
\end{figure}

Data in this section uses the qubit numbering ($Q_{FN}$ contrasting with $Q_{N}$ in Fig.~\ref{fig:RoundSched}) notation presented in Fig.~\ref{fig:Fnum_ZZa}, matching standard IBM Quantum Falcon systems. Summarized in Table ~\ref{table:1Q_Benchmark} are single qubit benchmarks for $ibm\_peekskill$, where single qubit gates for all qubits (excluding virtual Z gates) are identically 35.55ns.  While the Falcon layout has 27 qubits, for the d=3 circuits presented in this paper we only needed to use 23 of those qubits as shown in Fig.~\ref{fig:Fnum_ZZa}, excluding qubits $Q_{F0}$, $Q_{F6}$, $Q_{F20}$, and $Q_{F26}$. 

The always-on coupling between connected qubits on $ibm\_peekskill$ also results in undesirable static ZZ, plotted in Fig.~\ref{fig:Fnum_ZZb}, as a function of qubit-qubit detuning. To mitigate some of these effects, a simple $X_\pi$-$X_\pi$ dynamical decoupling sequence is added to code qubits throughout the circuit. Furthermore, by introducing mixed dimensionality simultaneous RB~\cite{McKay19_3QRB}, we can further capture the undesired side-effects of this coupling by comparing one and two-qubit gate error taken with standard RB with spectator qubits/gates fully idling or with those simultaneously driven as set by scheduling requirements of the $Z$ and $X$ checks. Simultaneous gate error for gates and qubits not part of these measurements (always idling during the experiments presented in the main text) are thus not included in this extra characterization (in table as NaN). These results are presented in Tables ~\ref{table:1Q_Benchmark} and ~\ref{table:2Q_Benchmark}. Optimization of two-qubit gates was undertaken on $ibm\_peekskill$ to ensure that no significant degradation in gate error or increase in leakage out of the computational manifold occurred in simultaneous benchmarking.

\begin{table}
\tabcolsep=0.2cm

\begin{tabular}{c c c c c c c c c c}
\toprule
 Qubit &  Freq. &  Anharm. &  $T_1$ &  $T_2$ & EPG &  EPG simul &  Readout  &  Initialization & Reset \\
 ($Q_F$) &  (GHz) &  (MHz) &  ($\mu$s) &  ($\mu$s) & (\%) &   (\%) & error (\%) & error (\%) & error (\%) \\
\hline
\midrule
    1 &        4.664 &         -351.7 &   420.3 &   118.4 & 0.0102 &     0.0143 &             1.24 &                    2.47 &                        3.3 \\
    2 &        4.799 &         -346.9 &   354.8 &   119.8 & 0.0128 &     0.0171 &             0.69 &                    1.27 &                        5.6 \\
    3 &        4.862 &         -347.9 &   331.7 &    25.8 & 0.0096 &        NaN &             0.99 &                    1.36 &                        5.4 \\
    4 &        4.933 &         -345.9 &   124.8 &    77.3 & 0.0332 &     0.0315 &             0.46 &                    0.52 &                        2.0 \\
    5 &        5.020 &         -343.9 &   131.7 &   215.5 & 0.0122 &     0.0145 &             0.84 &                    1.06 &                        1.2 \\
    7 &        4.769 &         -347.1 &   424.5 &    59.7 & 0.0107 &     0.0212 &             0.55 &                    0.31 &                        3.3 \\
    8 &        4.941 &         -344.3 &   249.4 &   228.8 & 0.0181 &     0.0310 &             0.48 &                    0.77 &                        1.1 \\
    9 &        5.219 &         -339.4 &   271.7 &   316.0 & 0.0069 &     0.0287 &             1.24 &                    1.95 &                        1.7 \\
    10 &        4.863 &         -347.1 &   357.0 &    72.0 & 0.0184 &     0.0207 &             0.26 &                    0.61 &                        1.8 \\
    11 &        5.128 &         -341.4 &   283.8 &   188.8 & 0.0199 &     0.0217 &             1.59 &                    2.69 &                        2.6 \\
    12 &        4.933 &         -344.8 &   280.9 &   353.0 & 0.0190 &     0.0367 &             0.36 &                    0.38 &                        1.2 \\
    13 &        5.006 &         -356.5 &   349.8 &   345.0 & 0.0168 &     0.0410 &             0.10 &                    0.48 &                        0.9 \\
    14 &        4.839 &         -377.2 &   399.3 &    99.7 & 0.0157 &     0.0694 &             1.19 &                    4.48 &                        4.6 \\
    15 &        4.991 &         -368.8 &   226.6 &   217.4 & 0.0352 &     0.0473 &             0.34 &                    1.38 &                        3.2 \\
    16 &        5.107 &         -342.0 &   259.8 &   209.2 & 0.0100 &     0.0280 &             0.75 &                    0.86 &                        1.2 \\
    17 &        5.173 &         -339.3 &   234.4 &   311.7 & 0.0207 &     0.0324 &             0.61 &                    1.43 &                        1.6 \\
    18 &        5.103 &         -339.9 &   195.5 &    34.7 & 0.0138 &     0.0118 &             0.28 &                    0.22 &                        1.2 \\
    19 &        4.819 &         -376.7 &   319.6 &   167.6 & 0.0311 &     0.0485 &             1.17 &                    3.28 &                        7.2 \\
    21 &        4.890 &         -345.8 &   278.1 &   308.0 & 0.0131 &     0.0143 &             0.50 &                    0.60 &                        0.7 \\
    22 &        4.955 &         -344.2 &   206.9 &   132.4 & 0.0105 &     0.0177 &             0.62 &                    0.90 &                        1.8 \\
    23 &        5.045 &         -341.8 &   278.3 &   145.0 & 0.0118 &        NaN &             0.23 &                    0.21 &                        0.8 \\
    24 &        5.136 &         -341.1 &   258.7 &    14.6 & 0.0147 &     0.0169 &             0.34 &                    0.58 &                        2.6 \\
    25 &        5.027 &         -341.7 &   364.1 &   327.5 & 0.0160 &     0.0265 &             0.40 &                    0.52 &                        0.8 \\
\bottomrule
\end{tabular}
\caption{Single qubit device parameters, using IBM-Falcon qubit numbering presented in Fig.~\ref{fig:Fnum_ZZa} for $ibm\_peekskill$. Single qubit error per gate (EPG) is obtained by performing randomized benchmarking (RB) on a given qubit with all coupled/spectator qubits idling. In contrast, simultaneous single qubit EPG (EPG simul) is obtained by performing one qubit RB concurrently two-qubit RB on neighboring gates (presented in Table \ref{table:2Q_Benchmark}). This is done to realistically approximate simultaneous application of gates in the $X$ and $Z$ stabilizers. To separate readout error from initialization and reset error, readout error is extracted from overlap of gaussian fits to ground and excited state histograms. The initialization sequence for the data presented in this paper was three rounds of conditional reset -  $X\pi_{e-f}$ - then three more rounds of conditional reset in an effort to reduce f state population while maintaining a fast experiment repetition rate.  The initialization error is that measured on average, after this sequence is applied simultaneously on all qubits.  Reset error is average non-zero state population after a single round of conditional reset (simultaneously done across all qubits) after preparing all qubits on with an $X_{\pi/2}$, to capture mid-circuit reset needed for each syndrome measurement round.}
\label{table:1Q_Benchmark}
\end{table}

\begin{table}
\tabcolsep=0.5cm
\begin{tabular}{cccc}
\toprule
 Gate &  CX length (ns) &      EPG (\%)&  EPG\_simul (\%)\\
\midrule
\hline
11\_14 &           483.6 & 0.53 &       0.95 \\
12\_15 &           433.8 & 0.78 &       1.42 \\
12\_10 &           334.2 & 0.42 &       1.02 \\
12\_13 &           519.1 & 0.61 &       1.03 \\
14\_13 &           504.9 & 0.77 &       1.49 \\
15\_18 &           469.3 & 0.49 &       1.10 \\
16\_14 &           440.9 & 0.50 &       0.82 \\
16\_19 &           696.9 & 2.09 &       1.14 \\
18\_17 &           426.7 & 4.03 &       3.96 \\
18\_21 &           348.4 & 0.56 &       0.73 \\
  2\_1 &           362.7 & 0.34 &       0.55 \\
21\_23 &           519.1 & 0.65 &       0.67 \\
22\_19 &           362.7 & 0.50 &       0.94 \\
22\_25 &           412.4 & 0.47 &       0.57 \\
24\_23 &           384.0 & 0.64 &       0.81 \\
24\_25 &           384.0 & 0.69 &       0.91 \\
  3\_2 &           426.7 & 0.52 &       0.53 \\
  3\_5 &           391.1 & 0.40 &       0.64 \\
  4\_1 &           547.6 & 0.46 &       0.49 \\
  5\_8 &           348.4 & 0.47 &       0.71 \\
 7\_10 &           362.7 & 1.42 &       0.76 \\
  7\_4 &           426.7 & 0.39 &       0.54 \\
 8\_11 &           526.2 & 1.16 &       1.39 \\
  9\_8 &           384.0 & 0.58 &       0.84 \\
\bottomrule
\end{tabular}
\caption{Two qubit gates used in $X$ and $Z$ stabilizers for $ibm\_peekskill$. The CX gates are constructed from the echoed cross-resonance gate \cite{sundaresanReducingUnitarySpectator2020}, with lengths and gate directions optimized for overall device performance. EPG is measured with spectator qubits idling while simultaneous EPG is taken with spectator qubits undergoing single qubit RB.}
\label{table:2Q_Benchmark}
\end{table}

Using the same methodology presented in \cite{chen2021calibrated}, reset operations conditioned on the preceding measurement result are used for mid-circuit reset operations shown Fig.~\ref{fig:RoundSched_b}. The total time of the measurement + reset cycle is 768ns, and includes an approximately 400ns  measurement pulse, cavity ring-down time overlapping with classical control path delays, and application of the conditional $X_\pi$. For consistency, all qubits are calibrated to use the same duration pulse and delays, with pulse amplitude calibrated individually to optimize QND-ness of readout. 

To optimize the performance of the analytical perfect matching decoding on experimental data, an optimization algorithm was run to find a set of input error parameters that minimizes the decoder output logical error rates. Here we chose to use the L-BFGS-B algorithm \cite{byrd1995limited} due to efficiency of optimization and ability to work with simple linear constraints. The optimization resulted in the following set of input error parameters for the analytical perfect matching decoding algorithm $p_C =$ [0.01, 0.0028, 0.0, 0.001, 0.002, 0.0028, 0.0028, 0.0, 0.0005, 0.0, 0.00001] following the error locations $C=\{\mathrm{cx},\mathrm{h},\mathrm{s},\mathrm{id},\mathrm{idm},\mathrm{x},\mathrm{y},\mathrm{z},\mathrm{measure},\mathrm{initialize}, \mathrm{reset} \}$ as defined in Section \ref{sec:matching}.

We use the following equation to fit logical errors at syndrome measurement round, $r$, 

\begin{equation}
     P_{\mathrm{fail}}(r) = \frac{1}{2}(1 - A e^{-r/\tau}) \label{eq:pfail_fit}
\end{equation}

where $A$ is SPAM error, $\tau=\frac{-1}{ln(1-2\epsilon)}$, and $\epsilon$ is the logical error rate per syndrome measurement round (~\ref{fig:RoundvsError_a}-inset and ~\ref{fig:RoundvsError_b})

\subsection{Leakage in the system}
\labelname{E}\label{leakage}
Leakage errors outside the computational space comprising the states $|0\rangle$ (\textit{g}-state) and $|1\rangle$ (\textit{e}-state) into $|2\rangle$ (\textit{f}-state) or higher states cannot be corrected by our quantum error correction code and thus pose a serious threat to fault-tolerant computing. For fixed-frequency superconducting qubits, a certain set of qubit frequency assignments may lead to a frequency \emph{collisions} during the cross-resonant gate operation~\cite{hertzberg2021laser}. For example, when the target qubit frequency is close to the $e\rightarrow f$ transition frequency of the control qubit, leakage error is induced during the two qubit gate operation. Another example is a simultaneous operation of a two-qubit gate with a spectator single-qubit gate where the spectator qubit frequency together with target qubit frequency match the $e\rightarrow f$ transition of the control qubit. This can result in leakage errors which can be characterized by randomized benchmarking of the corresponding single- and two-qubit gates~\cite{Wood2018Quantification}.         

\begin{figure}[h]
    \centering
    \begin{minipage}{.35\linewidth}
    \vspace*{0.5cm}
    \subfloat[]{{\includegraphics[height=0.06\textheight]{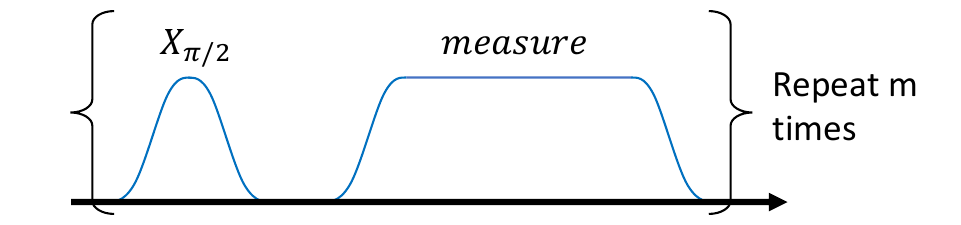}\label{fig:lkg_a}}} \\[-0.5ex]
    \subfloat[]{{\includegraphics[height=0.135\textheight]{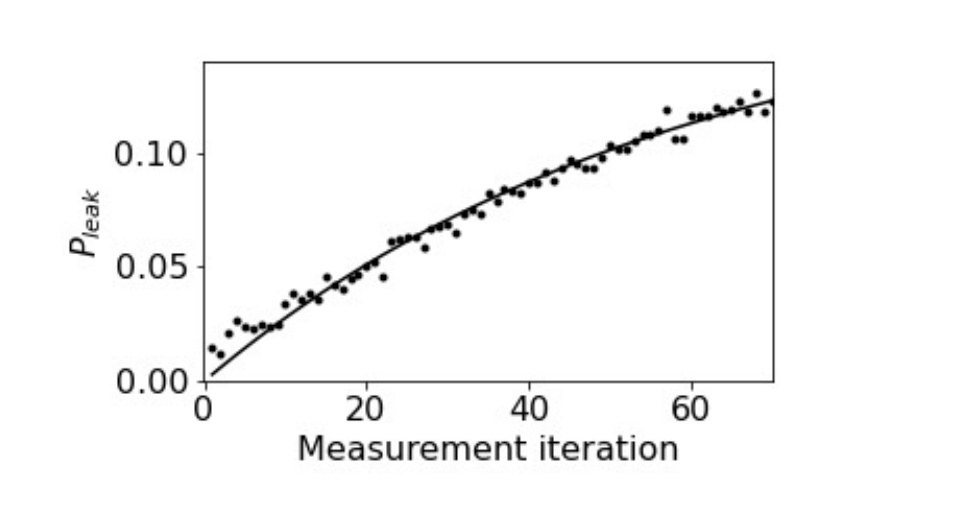}\label{fig:lkg_b}}}
    \end{minipage}
    \begin{minipage}{.6\linewidth}
    \subfloat[\centering]{{\includegraphics[height=0.22\textheight]{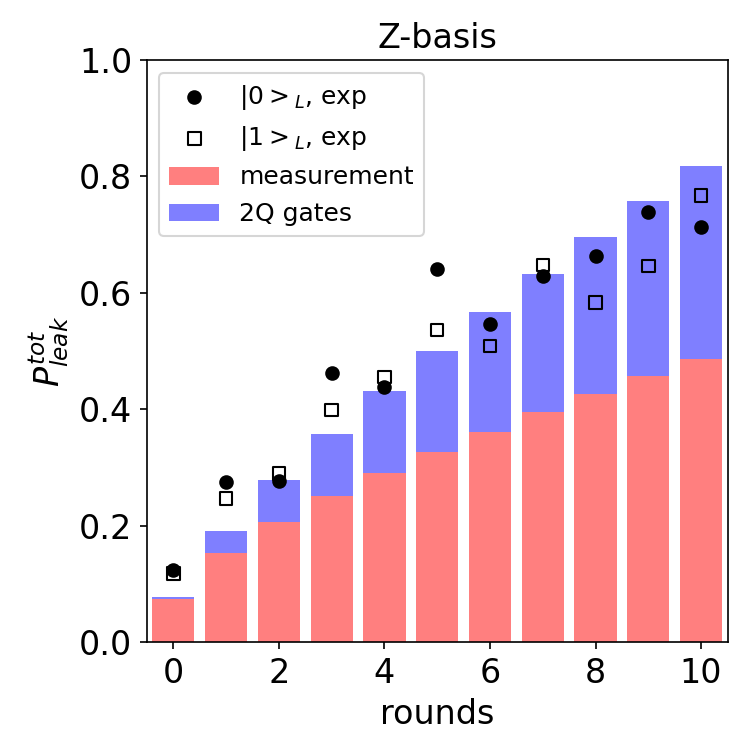}\label{fig:lkg_c}}}
    \subfloat[\centering]{{\includegraphics[height=0.22\textheight]{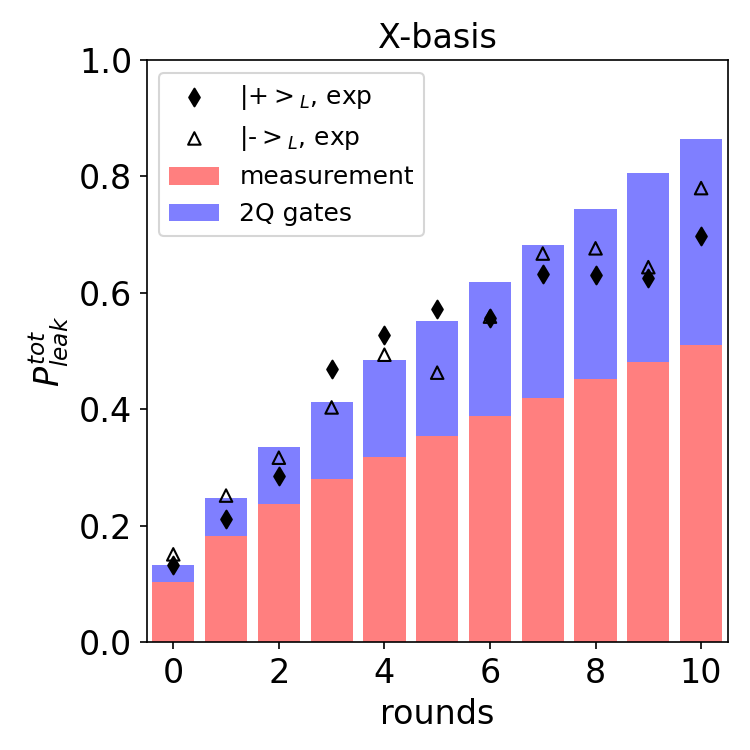}\label{fig:lkg_d}}}
    \end{minipage}
    \caption{(a) Repeated measurement sequence for extracting leakage error during the measurement. The $X_{\pi/2}$ pulse allows us to randomly sample leakage events from $|0\rangle$ or $|1\rangle$ states. (b) The leakage probability ($p_{\mathrm{leak}}^{\mathrm{meas}}$) to the $|2\rangle$ state measured at $Q_{F14}$. The leakage and seepage rate is obtained by fitting the data with Eq.~\ref{eq:lkg}. (c, d)  Qubit leakage in the system as a function of syndrome measurement rounds for $Z-$ and $X-$basis logical states. Bar plots show the $p^{\mathrm{tot}}_{\mathrm{leak}}$ as computed from the gate and measurement leakage rates, obtained from randomized benchmarking (2Q gates) and from the sequence shown in (a), respectively. Experimental results, $p_{\mathrm{leak}}^{\mathrm{exp}}=1-p_{\mathrm{accept}}$, where $p_{\mathrm{accept}}$ is the acceptance probability calculated from the method outlined in \hyperref[supp:post-select-method]{Supp.~\nameref{supp:post-select-method}}, are shown as black symbols for comparison. The experimental results plotted here do not include initialization leakage.}
    \label{fig:lkg}
\end{figure}

Leakage errors can also occur during measurements~\cite{Sank2016Measurement}. As we speed up the measurement time by increasing the measurement power, qubits become more prone to leakage. We characterize this measurement-induced leakage by repeatedly measuring the qubit and extracting the leakage rate. The experiment is described in Fig.~\ref{fig:lkg_a}, where the sequence consists of $X_{\pi/2}$ followed by a measurement tone. The $X_{\pi/2}$ pulse will map either $|0\rangle$ or $|1\rangle$ to the equator of the Bloch sphere, so the sequence randomly samples either $|0\rangle$ or $|1\rangle$ during the subsequent measurement. The obtained measurement leakage rate thus obtained is an average of the leakage rates from $|0\rangle$ and $|1\rangle$ states. The outcomes obtained from the sequence in Fig.~\ref{fig:lkg_a} are classified according to calibration data obtained by preparing the $|0\rangle$, $|1\rangle$, and $|2\rangle$ states, using the closest distribution mean for each outcome, and then applying readout error mitigation by constraining the formalism described in \cite{Bravyi2021Mitigating} for multi-qubit readout to our single-qubit three-state subspace. This single-qubit readout error mitigation is applied to the ensemble of measurements obtained for each iteration of the pulse sequence. The measurement sequence is repeated for $m=70$ times and we average over the $10,000$ shots for each $m$ to compute the averaged probability that the qubit is binned in the $|2\rangle$ state. Fig.~\ref{fig:lkg_b} shows the measurement leakage probability, $p_{\mathrm{leak}}^{\mathrm{meas}}$, where the qubit leaks to the $|2\rangle$ state per measurement. Eventually a steady state population in the $|2\rangle$ state, determined by the measurement leakage and seepage rates, is reached. We extract the leakage and seepage rates using the equation  
\begin{equation} \label{eq:lkg}
    p_{\mathrm{leak}}^{\mathrm{meas}}=\frac{\Gamma_L}{\Gamma_L+\Gamma_S}\left( 1-e^{-(\Gamma_L+\Gamma_S)m} \right),
\end{equation}
where the leakage rate $\Gamma_L$ is the probability of the qubit leaking during a measurement, the seepage rate $\Gamma_S$ is the probability of a leaked state returning to the qubit subspace during a measurement. Here, $\Gamma_{L,S}$ measures rate per measurement, therefore it is a unitless quantity. The obtained average and median value of $\Gamma_L$ are $6.54\times 10^{-3}$ and $4.86\times 10^{-3}$ per measurement, respectively. 

We extract the two-qubit gate leakage and seepage rate of the $|2\rangle$ state from simultaneous randomized benchmarking, with the simultaneity chosen to match the $Z-$ and $X-$stabilizer sequences as illustrated in Fig.~\ref{fig:RoundSched}. Similarly, we extract the leakage/seepage rate from repeated measurement described in Fig.~\ref{fig:lkg_a}. In these estimations, we account for the number of gate operations and measurements for each syndrome/flag qubits as well as the code qubits measured at the end. For instance, a two round experiment for the logical $Z-$basis consists of an $X-$check for state preparation, two rounds of $X-$ and $Z-$checks, and a final measurement of the code qubits. Each check consists of two-qubit gates and measurements. As a result, there are three sets of two-qubit gates and measurements on $X-$check qubits, two sets of two-qubit gates and measurements on $Z-$check qubits, and one measurement of the code qubits. The post-selection procedure discards the result if any of the qubit is leaked from the computational subspace. Therefore, we sum all the leakage probabilities to compute $p_{\mathrm{leak}}^{\mathrm{tot}}$ for each syndrome measurement round. Fig.~\ref{fig:lkg_c},~\ref{fig:lkg_d} shows $p_{\mathrm{leak}}^{\mathrm{tot}}$ as a function of the number of rounds for the $Z-$ and $X-$ logical bases, respectively. Each bar represents $p_{\mathrm{leak}}^{\mathrm{tot}}$ from two-qubit gates (blue) and measurement (red) operations. The leakage error caused by measurement contributes the most for early rounds, then tends to saturate. The leakage contribution from two-qubit gates becomes significant for later rounds. 

This analysis shows that reducing leakage error from both two-qubit gates and measurements is important. Decreasing leakage induced by two-qubit gates in our architecture will be associated with slower gates. With respect to measurement, as noted above, it is well known that a strong drive on a superconducting qubit system can lead to transitions both beyond the computational space~\cite{Sank2016Measurement} and beyond the confinement of the Josephson cosine potential~\cite{Lescanne19}. There is therefore a trade-off to be considered between readout error and measurement length and leakage probability. Slower readout impacts the system by increasing the idle time of the qubits not being measured. There have been proposals to deal with leakage in superconducting qubit systems by moving all the qubit excitations to the readout resonator, from which they decay to the environment~\cite{McEwen21}, or by designing readout resonator leakage reduction units (LRU)~\cite{Battistel21} which exploit particular transition levels of the qubit-resonator system and which transform leakage errors into Pauli errors. LRU have also been proposed at the code level~\cite{SucharaLRU}. These options, as well as higher branching capabilities in readout and control electronics to conditionally reset qubits to the ground state from higher excitation levels, could be explored in experimental systems demonstrating quantum error correction in the near future.

\subsection{Post-selection method}
\begin{figure}[h]
	\centering
	\subfloat[\centering]{{\includegraphics[height=0.2\textheight]{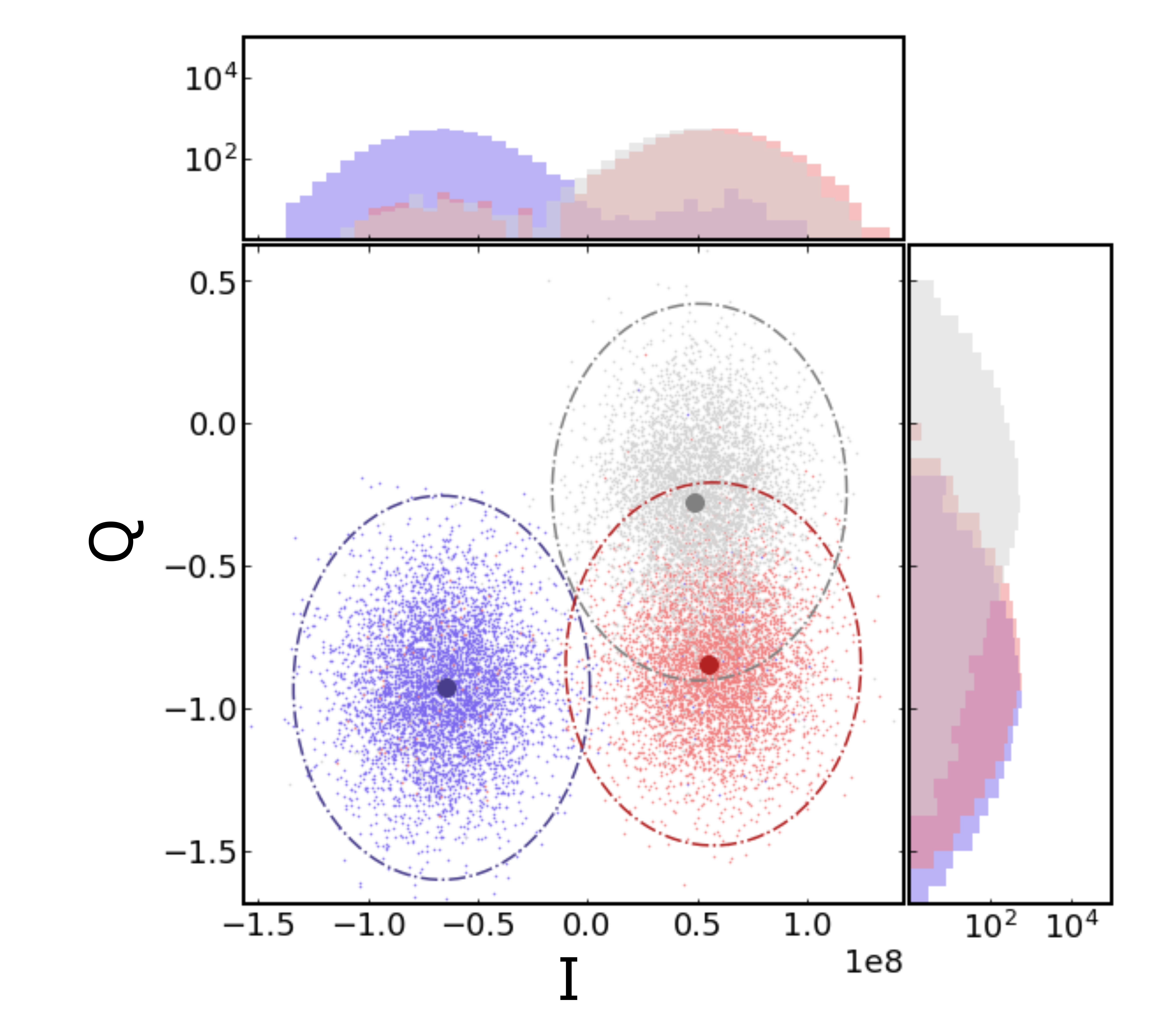}\label{fig:post_select_a}}}
	\qquad
	\subfloat[\centering]{{\includegraphics[height=0.2\textheight]{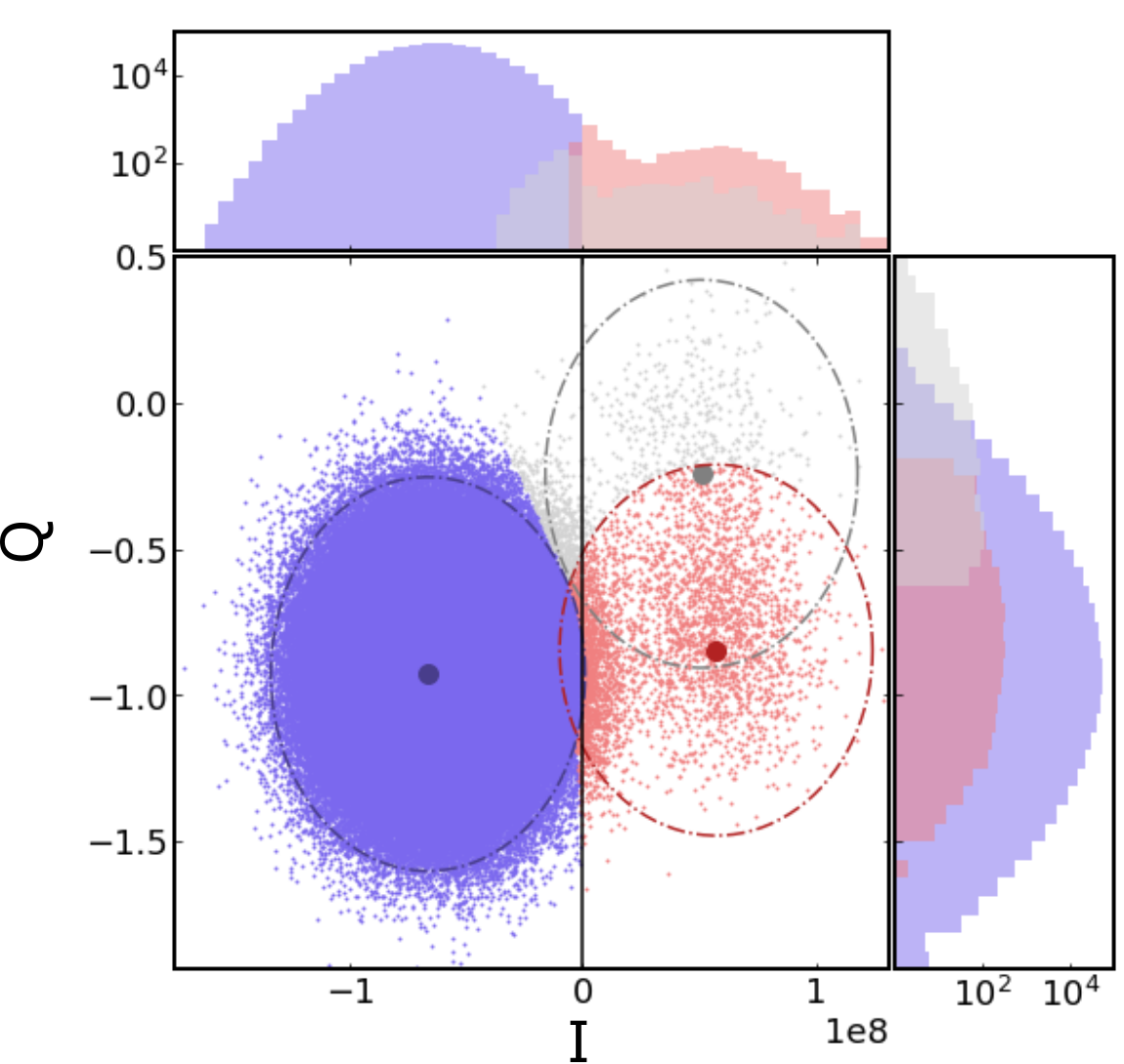}\label{fig:post_select_b}}}
	\qquad
	\subfloat[\centering]{{\includegraphics[height=0.2\textheight]{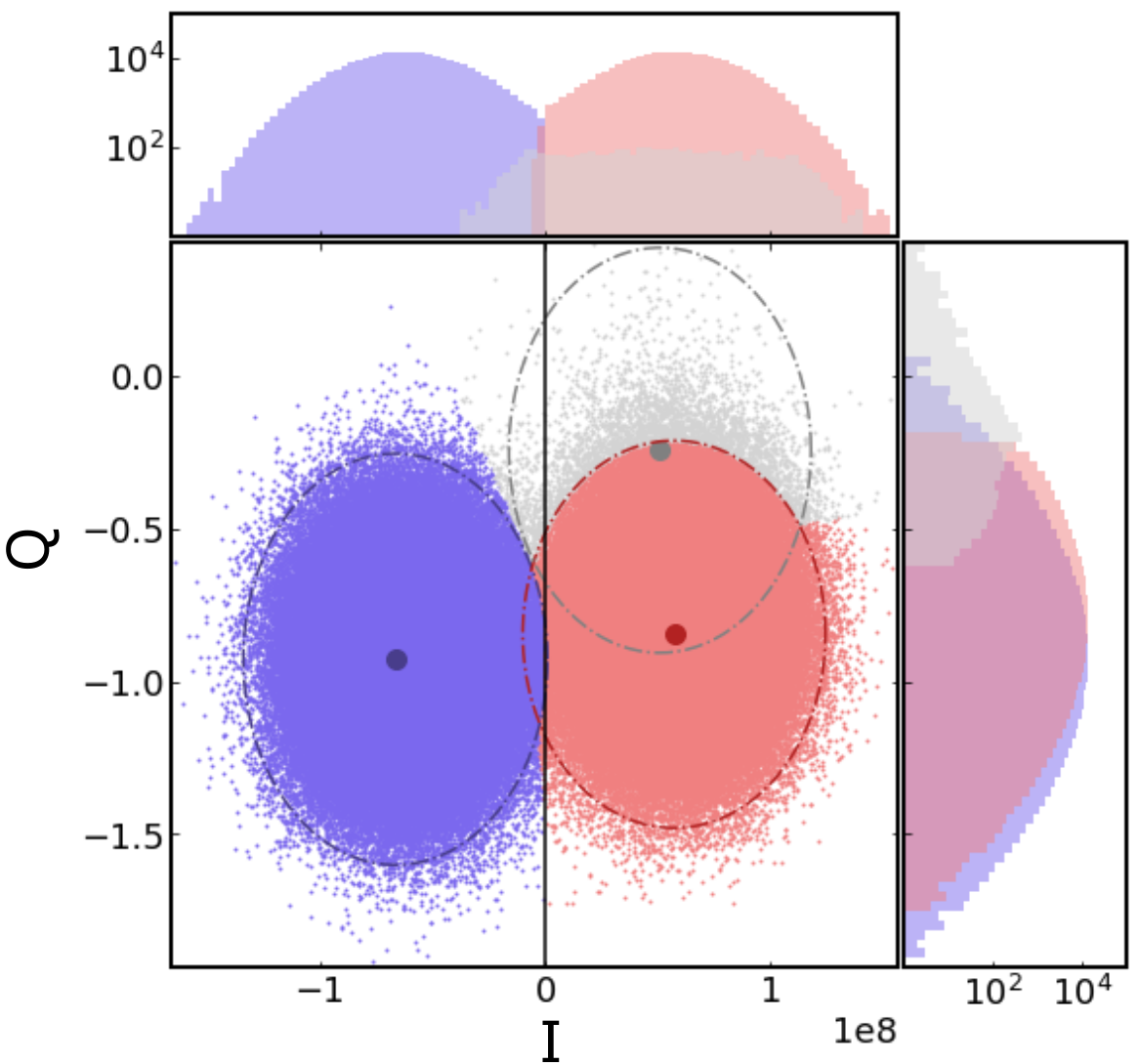}\label{fig:post_select_c}}}
			\caption{(a) Readout calibration data for $Q_{F12}$ (see Fig.~\ref{fig:Fnum_ZZa}). The qubit is prepared in its $|0\rangle$, $|1\rangle$, and $|2\rangle$ states and measured. The collected statistics can be seen in as blue ($|0\rangle$), red ($|1\rangle$), and grey ($|2\rangle$) where the dot-dashed lines represent 3-$\sigma$ for each distribution. (b) 3-state classification results for $Q_{F12}$ after qubit initialization, and (c) after the first $X-$syndrome measurement.}
	\label{fig:post_select}
\end{figure}

\labelname{F}\label{supp:post-select-method}
We post-select all our results to remove detected leakage events in any of the qubits in our system. To do this, we look at 5,000 integrated outputs for each qubit when prepared in each of the states $|0\rangle$, $|1\rangle$, and $|2\rangle$. We show this calibration for $Q_{F12}$ (see Fig~\ref{fig:Fnum_ZZa}) in Fig.~\ref{fig:post_select_a}. The overlap between the $|1\rangle$ and $|2\rangle$ states, which is significant in all 23 qubits used in this work, makes the classification of these states challenging. Furthermore, the presence of decay events ($|1\rangle$ to $|0\rangle$, $|2\rangle$ to $|1\rangle$, or $|2\rangle$ to $|0\rangle$) may impair the results using this training data within a supervised learning protocol. We instead apply clustering methods to our calibration data using a Gaussian Mixture Model (GMM) with three clusters, each cluster with an independent diagonal covariance matrix. The diagonal entries of the covariance matrices can be used to extract the standard deviations of the distribution for each qubit state. This offers a convenient way for us to define more flexible classification rules, compared to, for example, simpler clustering algorithms like K-means. Once the centroids and standard deviations ($\sigma_x$ and $\sigma_y$) are determined from the calibration data, we define regions for each state within the \textit{I}/\textit{Q} plane determined by a radius of $3\sigma$ on each axis around the corresponding centroid (see Fig.~\ref{fig:post_select}). 

For any given measurement in any of the qubits, if the integrated outcome is within the $|0\rangle$-state region \textit{and} the $I$-quadrature is negative, we classify that outcome as $|0\rangle$. If the integrated outcome is not within the $|0\rangle$-state region \textit{or} the $I$-quadrature is positive, if it is within the $|1\rangle$-state region we classify it as $|1\rangle$, and if it is within the $|2\rangle$-state region but not within the $|1\rangle$-state region, we classify it as $|2\rangle$. For all other results, we classify the output according to its closest centroid. 

This classification method is applied to every qubit after every measurement and the experimental runs in which any qubit is measured as $|2\rangle$ is discarded. Fig.~\ref{fig:post_select_b} shows the readout outcomes of $Q_{F12}$ after the last initialization measurement. We only discard uncorrectable errors ($|2\rangle$ state) and retain experimental shots in which a qubit is in the $|1\rangle$ state after initialization, as that should be a correctable error by the code. Fig.~\ref{fig:post_select_c} shows the $Q_{F12}$ results after the first $X-$check for a logical $|0\rangle$ state preparation. Both the initialization and the mid-circuit contain the 500,000 shots that are used for each error correction run in our experiments. For the initialization classification we obtain populations of 0.9910, 0.0071, and 0.0019 for the $|0\rangle$, $|1\rangle$, and $|2\rangle$ states, respectively. For the mid-circuit $X-$syndrome classification, those populations are observed to be 0.4972, 0.4962, and 0.0066.

\subsection{Error for $r=2$ rounds}\labelname{G}\label{supp:neq2}
Table~\ref{table:Decoders} shows a comparison across the decoders studied in this work for state preparation and two rounds of syndrome measurement for the logical states $|+\rangle_L$, $|-\rangle_L$, $|0\rangle_L$, and $|1\rangle_L$. The results for the matching decoder with analytical input for the states $|0\rangle_L$ and $|+\rangle_L$ correspond to the values shown in Fig.~\ref{fig:RoundvsError_a} for $r=2$ rounds. 
\begin{table}[h]
\begin{tabular}{|c|c|c||p{1.5cm}|p{1.5cm}|p{1.5cm}|p{1.5cm}|p{1.5cm}|p{1.5cm}|p{1.5cm}|}
\hline
Basis & Init. State & Round Schedule & Matching Uniform (Full) & Matching Uniform (PS) & Matching Analytical (Full) & Matching Analytic (PS) & Maximum likelihood (Full) & Maximum likelihood (PS) & Shots (PS) \\
\hline
X & $\vert+\rangle_L$ & ZXZXZ  & 0.2555(6) & 0.2212(7) & 0.2502(6)  & 0.2091(7) & 0.2502(6) & 0.2083(7) & 317,672 \\
X & $\vert-\rangle_L$ & ZXZXZ  & 0.2860(6) & 0.2468(8) & 0.2805(6)   & 0.2332(8) & 0.2803(6) & 0.2321(8) & 295,608 \\
Z & $\vert0\rangle_L$ & XZXZX  & 0.1187(5) & 0.0978(5) & 0.1160(5)   & 0.0940(5) & 0.1045(4) & 0.0843(5) & 322,165 \\
Z & $\vert1\rangle_L$ & XZXZX  & 0.1151(5) & 0.0928(5) & 0.1162(5)   & 0.0920(5) & 0.1031(4) & 0.0819(5) & 306,962 \\
\hline
\end{tabular}
\caption{Comparison of logical error extracted using matching uniform, matching analytical, and maximum likelihood decoders on both full and leakage-post selected (PS) data-sets for $r=2$ rounds. The uncertainty corresponds to sampling noise, with each full data set corresponding to 500,000 shots, and the post-selected data sets keep a number of shots shown in the last column.}
\label{table:Decoders}
\end{table}


\begin{thebibliography}{43}%
	\makeatletter
	\providecommand \@ifxundefined [1]{%
		\@ifx{#1\undefined}
	}%
	\providecommand \@ifnum [1]{%
		\ifnum #1\expandafter \@firstoftwo
		\else \expandafter \@secondoftwo
		\fi
	}%
	\providecommand \@ifx [1]{%
		\ifx #1\expandafter \@firstoftwo
		\else \expandafter \@secondoftwo
		\fi
	}%
	\providecommand \natexlab [1]{#1}%
	\providecommand \enquote  [1]{``#1''}%
	\providecommand \bibnamefont  [1]{#1}%
	\providecommand \bibfnamefont [1]{#1}%
	\providecommand \citenamefont [1]{#1}%
	\providecommand \href@noop [0]{\@secondoftwo}%
	\providecommand \href [0]{\begingroup \@sanitize@url \@href}%
	\providecommand \@href[1]{\@@startlink{#1}\@@href}%
	\providecommand \@@href[1]{\endgroup#1\@@endlink}%
	\providecommand \@sanitize@url [0]{\catcode `\\12\catcode `\$12\catcode
		`\&12\catcode `\#12\catcode `\^12\catcode `\_12\catcode `\%12\relax}%
	\providecommand \@@startlink[1]{}%
	\providecommand \@@endlink[0]{}%
	\providecommand \url  [0]{\begingroup\@sanitize@url \@url }%
	\providecommand \@url [1]{\endgroup\@href {#1}{\urlprefix }}%
	\providecommand \urlprefix  [0]{URL }%
	\providecommand \Eprint [0]{\href }%
	\providecommand \doibase [0]{https://doi.org/}%
	\providecommand \selectlanguage [0]{\@gobble}%
	\providecommand \bibinfo  [0]{\@secondoftwo}%
	\providecommand \bibfield  [0]{\@secondoftwo}%
	\providecommand \translation [1]{[#1]}%
	\providecommand \BibitemOpen [0]{}%
	\providecommand \bibitemStop [0]{}%
	\providecommand \bibitemNoStop [0]{.\EOS\space}%
	\providecommand \EOS [0]{\spacefactor3000\relax}%
	\providecommand \BibitemShut  [1]{\csname bibitem#1\endcsname}%
	\let\auto@bib@innerbib\@empty
	\bibitem [{\citenamefont {Chamberland}\ \emph {et~al.}(2020)\citenamefont
		{Chamberland}, \citenamefont {Zhu}, \citenamefont {Yoder}, \citenamefont
		{Hertzberg},\ and\ \citenamefont
		{Cross}}]{chamberlandTopologicalSubsystemCodes2020}%
	\BibitemOpen
	\bibfield  {author} {\bibinfo {author} {\bibfnamefont {C.}~\bibnamefont
			{Chamberland}}, \bibinfo {author} {\bibfnamefont {G.}~\bibnamefont {Zhu}},
		\bibinfo {author} {\bibfnamefont {T.~J.}\ \bibnamefont {Yoder}}, \bibinfo
		{author} {\bibfnamefont {J.~B.}\ \bibnamefont {Hertzberg}},\ and\ \bibinfo
		{author} {\bibfnamefont {A.~W.}\ \bibnamefont {Cross}},\ }\bibfield  {title}
	{\bibinfo {title} {Topological and subsystem codes on low-degree graphs with
			flag qubits},\ }\href {https://doi.org/10.1103/PhysRevX.10.011022} {\bibfield
		{journal} {\bibinfo  {journal} {Physical Review X}\ }\textbf {\bibinfo
			{volume} {10}},\ \bibinfo {pages} {011022} (\bibinfo {year} {2020})},\
	\Eprint {https://arxiv.org/abs/1907.09528} {arXiv:1907.09528} \BibitemShut
	{NoStop}%
	\bibitem [{\citenamefont {Hertzberg}\ \emph {et~al.}(2021)\citenamefont
		{Hertzberg}, \citenamefont {Zhang}, \citenamefont {Rosenblatt}, \citenamefont
		{Magesan}, \citenamefont {Smolin}, \citenamefont {Yau}, \citenamefont
		{Adiga}, \citenamefont {Sandberg}, \citenamefont {Brink}, \citenamefont
		{Chow} \emph {et~al.}}]{hertzberg2021laser}%
	\BibitemOpen
	\bibfield  {author} {\bibinfo {author} {\bibfnamefont {J.~B.}\ \bibnamefont
			{Hertzberg}}, \bibinfo {author} {\bibfnamefont {E.~J.}\ \bibnamefont
			{Zhang}}, \bibinfo {author} {\bibfnamefont {S.}~\bibnamefont {Rosenblatt}},
		\bibinfo {author} {\bibfnamefont {E.}~\bibnamefont {Magesan}}, \bibinfo
		{author} {\bibfnamefont {J.~A.}\ \bibnamefont {Smolin}}, \bibinfo {author}
		{\bibfnamefont {J.-B.}\ \bibnamefont {Yau}}, \bibinfo {author} {\bibfnamefont
			{V.~P.}\ \bibnamefont {Adiga}}, \bibinfo {author} {\bibfnamefont
			{M.}~\bibnamefont {Sandberg}}, \bibinfo {author} {\bibfnamefont
			{M.}~\bibnamefont {Brink}}, \bibinfo {author} {\bibfnamefont {J.~M.}\
			\bibnamefont {Chow}}, \emph {et~al.},\ }\bibfield  {title} {\bibinfo {title}
		{Laser-annealing josephson junctions for yielding scaled-up superconducting
			quantum processors},\ }\href@noop {} {\bibfield  {journal} {\bibinfo
			{journal} {npj Quantum Information}\ }\textbf {\bibinfo {volume} {7}},\
		\bibinfo {pages} {1} (\bibinfo {year} {2021})}\BibitemShut {NoStop}%
	\bibitem [{\citenamefont {Poulin}(2005)}]{Poulin2005Subsystem}%
	\BibitemOpen
	\bibfield  {author} {\bibinfo {author} {\bibfnamefont {D.}~\bibnamefont
			{Poulin}},\ }\bibfield  {title} {\bibinfo {title} {Stabilizer formalism for
			operator quantum error correction},\ }\href@noop {} {\bibfield  {journal}
		{\bibinfo  {journal} {Phys. Rev. Lett.}\ }\textbf {\bibinfo {volume} {95}}
		(\bibinfo {year} {2005})}\BibitemShut {NoStop}%
	\bibitem [{\citenamefont {Dennis}\ \emph {et~al.}(2002)\citenamefont {Dennis},
		\citenamefont {Kitaev}, \citenamefont {Landahl},\ and\ \citenamefont
		{Preskill}}]{Dennis2002TopologicalQM}%
	\BibitemOpen
	\bibfield  {author} {\bibinfo {author} {\bibfnamefont {E.}~\bibnamefont
			{Dennis}}, \bibinfo {author} {\bibfnamefont {A.~Y.}\ \bibnamefont {Kitaev}},
		\bibinfo {author} {\bibfnamefont {A.~J.}\ \bibnamefont {Landahl}},\ and\
		\bibinfo {author} {\bibfnamefont {J.}~\bibnamefont {Preskill}},\ }\bibfield
	{title} {\bibinfo {title} {Topological quantum memory},\ }\href@noop {}
	{\bibfield  {journal} {\bibinfo  {journal} {Journal of Mathematical Physics}\
		}\textbf {\bibinfo {volume} {43}},\ \bibinfo {pages} {4452} (\bibinfo {year}
		{2002})}\BibitemShut {NoStop}%
	\bibitem [{\citenamefont {Bombin}\ and\ \citenamefont
		{Martin-Delgado}(2006)}]{Bombin06Color}%
	\BibitemOpen
	\bibfield  {author} {\bibinfo {author} {\bibfnamefont {H.}~\bibnamefont
			{Bombin}}\ and\ \bibinfo {author} {\bibfnamefont {M.~A.}\ \bibnamefont
			{Martin-Delgado}},\ }\bibfield  {title} {\bibinfo {title} {Topological
			quantum distillation},\ }\href
	{https://doi.org/10.1103/PhysRevLett.97.180501} {\bibfield  {journal}
		{\bibinfo  {journal} {Phys. Rev. Lett.}\ }\textbf {\bibinfo {volume} {97}},\
		\bibinfo {pages} {180501} (\bibinfo {year} {2006})}\BibitemShut {NoStop}%
	\bibitem [{\citenamefont {Piveteau}\ \emph {et~al.}(2021)\citenamefont
		{Piveteau}, \citenamefont {Sutter}, \citenamefont {Bravyi}, \citenamefont
		{Gambetta},\ and\ \citenamefont {Temme}}]{Piveteau21}%
	\BibitemOpen
	\bibfield  {author} {\bibinfo {author} {\bibfnamefont {C.}~\bibnamefont
			{Piveteau}}, \bibinfo {author} {\bibfnamefont {D.}~\bibnamefont {Sutter}},
		\bibinfo {author} {\bibfnamefont {S.}~\bibnamefont {Bravyi}}, \bibinfo
		{author} {\bibfnamefont {J.~M.}\ \bibnamefont {Gambetta}},\ and\ \bibinfo
		{author} {\bibfnamefont {K.}~\bibnamefont {Temme}},\ }\bibfield  {title}
	{\bibinfo {title} {Error mitigation for universal gates on encoded qubits},\
	}\href {https://doi.org/10.1103/PhysRevLett.127.200505} {\bibfield  {journal}
		{\bibinfo  {journal} {Phys. Rev. Lett.}\ }\textbf {\bibinfo {volume} {127}},\
		\bibinfo {pages} {200505} (\bibinfo {year} {2021})}\BibitemShut {NoStop}%
	\bibitem [{\citenamefont {Chamberland}\ \emph {et~al.}(2018)\citenamefont
		{Chamberland}, \citenamefont {Iyer},\ and\ \citenamefont
		{Poulin}}]{Chamberland2018faulttolerant}%
	\BibitemOpen
	\bibfield  {author} {\bibinfo {author} {\bibfnamefont {C.}~\bibnamefont
			{Chamberland}}, \bibinfo {author} {\bibfnamefont {P.}~\bibnamefont {Iyer}},\
		and\ \bibinfo {author} {\bibfnamefont {D.}~\bibnamefont {Poulin}},\
	}\bibfield  {title} {\bibinfo {title} {Fault-tolerant quantum computing in
			the {P}auli or {C}lifford frame with slow error diagnostics},\ }\href
	{https://doi.org/10.22331/q-2018-01-04-43} {\bibfield  {journal} {\bibinfo
			{journal} {{Quantum}}\ }\textbf {\bibinfo {volume} {2}},\ \bibinfo {pages}
		{43} (\bibinfo {year} {2018})}\BibitemShut {NoStop}%
	\bibitem [{\citenamefont {DiVincenzo}\ and\ \citenamefont
		{Aliferis}(2007)}]{DiVincenzo07}%
	\BibitemOpen
	\bibfield  {author} {\bibinfo {author} {\bibfnamefont {D.~P.}\ \bibnamefont
			{DiVincenzo}}\ and\ \bibinfo {author} {\bibfnamefont {P.}~\bibnamefont
			{Aliferis}},\ }\bibfield  {title} {\bibinfo {title} {Effective fault-tolerant
			quantum computation with slow measurements},\ }\href
	{https://doi.org/10.1103/PhysRevLett.98.020501} {\bibfield  {journal}
		{\bibinfo  {journal} {Phys. Rev. Lett.}\ }\textbf {\bibinfo {volume} {98}},\
		\bibinfo {pages} {020501} (\bibinfo {year} {2007})}\BibitemShut {NoStop}%
	\bibitem [{\citenamefont {Linke}\ \emph {et~al.}()\citenamefont {Linke},
		\citenamefont {Gutierrez}, \citenamefont {Landsman}, \citenamefont {Figgatt},
		\citenamefont {Debnath}, \citenamefont {Brown},\ and\ \citenamefont
		{Monroe}}]{linkeFaulttolerantQuantumError}%
	\BibitemOpen
	\bibfield  {author} {\bibinfo {author} {\bibfnamefont {N.~M.}\ \bibnamefont
			{Linke}}, \bibinfo {author} {\bibfnamefont {M.}~\bibnamefont {Gutierrez}},
		\bibinfo {author} {\bibfnamefont {K.~A.}\ \bibnamefont {Landsman}}, \bibinfo
		{author} {\bibfnamefont {C.}~\bibnamefont {Figgatt}}, \bibinfo {author}
		{\bibfnamefont {S.}~\bibnamefont {Debnath}}, \bibinfo {author} {\bibfnamefont
			{K.~R.}\ \bibnamefont {Brown}},\ and\ \bibinfo {author} {\bibfnamefont
			{C.}~\bibnamefont {Monroe}},\ }\bibfield  {title} {\bibinfo {title}
		{Fault-tolerant quantum error detection},\ }\href
	{https://doi.org/10.1126/sciadv.1701074} {\bibfield  {journal} {\bibinfo
			{journal} {Science Advances}\ }\textbf {\bibinfo {volume} {3}},\ \bibinfo
		{pages} {e1701074}}\BibitemShut {NoStop}%
	\bibitem [{\citenamefont {Abobeih}\ \emph {et~al.}(2021)\citenamefont
		{Abobeih}, \citenamefont {Wang}, \citenamefont {Randall}, \citenamefont
		{Loenen}, \citenamefont {Bradley}, \citenamefont {Markham}, \citenamefont
		{Twitchen}, \citenamefont {Terhal},\ and\ \citenamefont
		{Taminiau}}]{abobeihFaulttolerantOperationLogical2021}%
	\BibitemOpen
	\bibfield  {author} {\bibinfo {author} {\bibfnamefont {M.~H.}\ \bibnamefont
			{Abobeih}}, \bibinfo {author} {\bibfnamefont {Y.}~\bibnamefont {Wang}},
		\bibinfo {author} {\bibfnamefont {J.}~\bibnamefont {Randall}}, \bibinfo
		{author} {\bibfnamefont {S.~J.~H.}\ \bibnamefont {Loenen}}, \bibinfo {author}
		{\bibfnamefont {C.~E.}\ \bibnamefont {Bradley}}, \bibinfo {author}
		{\bibfnamefont {M.}~\bibnamefont {Markham}}, \bibinfo {author} {\bibfnamefont
			{D.~J.}\ \bibnamefont {Twitchen}}, \bibinfo {author} {\bibfnamefont {B.~M.}\
			\bibnamefont {Terhal}},\ and\ \bibinfo {author} {\bibfnamefont {T.~H.}\
			\bibnamefont {Taminiau}},\ }\bibfield  {title} {\bibinfo {title}
		{Fault-tolerant operation of a logical qubit in a diamond quantum
			processor},\ }\href@noop {} {\bibfield  {journal} {\bibinfo  {journal}
			{arXiv:2108.01646 [cond-mat, physics:quant-ph]}\ } (\bibinfo {year}
		{2021})},\ \Eprint {https://arxiv.org/abs/2108.01646} {arXiv:2108.01646
		[cond-mat, physics:quant-ph]} \BibitemShut {NoStop}%
	\bibitem [{\citenamefont {Takita}\ \emph {et~al.}(2017)\citenamefont {Takita},
		\citenamefont {Cross}, \citenamefont {C{\'o}rcoles}, \citenamefont {Chow},\
		and\ \citenamefont
		{Gambetta}}]{takitaExperimentalDemonstrationFaulttolerant2017}%
	\BibitemOpen
	\bibfield  {author} {\bibinfo {author} {\bibfnamefont {M.}~\bibnamefont
			{Takita}}, \bibinfo {author} {\bibfnamefont {A.~W.}\ \bibnamefont {Cross}},
		\bibinfo {author} {\bibfnamefont {A.~D.}\ \bibnamefont {C{\'o}rcoles}},
		\bibinfo {author} {\bibfnamefont {J.~M.}\ \bibnamefont {Chow}},\ and\
		\bibinfo {author} {\bibfnamefont {J.~M.}\ \bibnamefont {Gambetta}},\
	}\bibfield  {title} {\bibinfo {title} {Experimental demonstration of
			fault-tolerant state preparation with superconducting qubits},\ }\href
	{https://doi.org/10.1103/PhysRevLett.119.180501} {\bibfield  {journal}
		{\bibinfo  {journal} {Physical review letters}\ }\textbf {\bibinfo {volume}
			{119}},\ \bibinfo {pages} {180501} (\bibinfo {year} {2017})}\BibitemShut
	{NoStop}%
	\bibitem [{\citenamefont {Andersen}\ \emph {et~al.}(2020)\citenamefont
		{Andersen}, \citenamefont {Remm}, \citenamefont {Lazar}, \citenamefont
		{Krinner}, \citenamefont {Lacroix}, \citenamefont {Norris}, \citenamefont
		{Gabureac}, \citenamefont {Eichler},\ and\ \citenamefont
		{Wallraff}}]{andersenRepeatedQuantumError2020a}%
	\BibitemOpen
	\bibfield  {author} {\bibinfo {author} {\bibfnamefont {C.~K.}\ \bibnamefont
			{Andersen}}, \bibinfo {author} {\bibfnamefont {A.}~\bibnamefont {Remm}},
		\bibinfo {author} {\bibfnamefont {S.}~\bibnamefont {Lazar}}, \bibinfo
		{author} {\bibfnamefont {S.}~\bibnamefont {Krinner}}, \bibinfo {author}
		{\bibfnamefont {N.}~\bibnamefont {Lacroix}}, \bibinfo {author} {\bibfnamefont
			{G.~J.}\ \bibnamefont {Norris}}, \bibinfo {author} {\bibfnamefont
			{M.}~\bibnamefont {Gabureac}}, \bibinfo {author} {\bibfnamefont
			{C.}~\bibnamefont {Eichler}},\ and\ \bibinfo {author} {\bibfnamefont
			{A.}~\bibnamefont {Wallraff}},\ }\bibfield  {title} {\bibinfo {title}
		{Repeated quantum error detection in a surface code},\ }\href
	{https://doi.org/10.1038/s41567-020-0920-y} {\bibfield  {journal} {\bibinfo
			{journal} {Nature Physics}\ }\textbf {\bibinfo {volume} {16}},\ \bibinfo
		{pages} {875} (\bibinfo {year} {2020})}\BibitemShut {NoStop}%
	\bibitem [{\citenamefont {Chen}\ \emph
		{et~al.}(2021{\natexlab{a}})\citenamefont {Chen}, \citenamefont {Satzinger},
		\citenamefont {Atalaya}, \citenamefont {Korotkov}, \citenamefont {Dunsworth},
		\citenamefont {Sank}, \citenamefont {Quintana}, \citenamefont {McEwen},
		\citenamefont {Barends}, \citenamefont {Klimov}, \citenamefont {Hong},
		\citenamefont {Jones}, \citenamefont {Petukhov}, \citenamefont {Kafri},
		\citenamefont {Demura}, \citenamefont {Burkett}, \citenamefont {Gidney},
		\citenamefont {Fowler}, \citenamefont {Paler}, \citenamefont {Putterman},
		\citenamefont {Aleiner}, \citenamefont {Arute}, \citenamefont {Arya},
		\citenamefont {Babbush}, \citenamefont {Bardin}, \citenamefont {Bengtsson},
		\citenamefont {Bourassa}, \citenamefont {Broughton}, \citenamefont {Buckley},
		\citenamefont {Buell}, \citenamefont {Bushnell}, \citenamefont {Chiaro},
		\citenamefont {Collins}, \citenamefont {Courtney}, \citenamefont {Derk},
		\citenamefont {Eppens}, \citenamefont {Erickson}, \citenamefont {Farhi},
		\citenamefont {Foxen}, \citenamefont {Giustina}, \citenamefont {Greene},
		\citenamefont {Gross}, \citenamefont {Harrigan}, \citenamefont {Harrington},
		\citenamefont {Hilton}, \citenamefont {Ho}, \citenamefont {Huang},
		\citenamefont {Huggins}, \citenamefont {Ioffe}, \citenamefont {Isakov},
		\citenamefont {Jeffrey}, \citenamefont {Jiang}, \citenamefont {Kechedzhi},
		\citenamefont {Kim}, \citenamefont {Kitaev}, \citenamefont {Kostritsa},
		\citenamefont {Landhuis}, \citenamefont {Laptev}, \citenamefont {Lucero},
		\citenamefont {Martin}, \citenamefont {McClean}, \citenamefont {McCourt},
		\citenamefont {Mi}, \citenamefont {Miao}, \citenamefont {Mohseni},
		\citenamefont {Montazeri}, \citenamefont {Mruczkiewicz}, \citenamefont
		{Mutus}, \citenamefont {Naaman}, \citenamefont {Neeley}, \citenamefont
		{Neill}, \citenamefont {Newman}, \citenamefont {Niu}, \citenamefont
		{O'Brien}, \citenamefont {Opremcak}, \citenamefont {Ostby}, \citenamefont
		{Pat{\'o}}, \citenamefont {Redd}, \citenamefont {Roushan}, \citenamefont
		{Rubin}, \citenamefont {Shvarts}, \citenamefont {Strain}, \citenamefont
		{Szalay}, \citenamefont {Trevithick}, \citenamefont {Villalonga},
		\citenamefont {White}, \citenamefont {Yao}, \citenamefont {Yeh},
		\citenamefont {Yoo}, \citenamefont {Zalcman}, \citenamefont {Neven},
		\citenamefont {Boixo}, \citenamefont {Smelyanskiy}, \citenamefont {Chen},
		\citenamefont {Megrant}, \citenamefont {Kelly},\ and\ \citenamefont {{Google
				Quantum AI}}}]{chenExponentialSuppressionBit2021}%
	\BibitemOpen
	\bibfield  {author} {\bibinfo {author} {\bibfnamefont {Z.}~\bibnamefont
			{Chen}}, \bibinfo {author} {\bibfnamefont {K.~J.}\ \bibnamefont {Satzinger}},
		\bibinfo {author} {\bibfnamefont {J.}~\bibnamefont {Atalaya}}, \bibinfo
		{author} {\bibfnamefont {A.~N.}\ \bibnamefont {Korotkov}}, \bibinfo {author}
		{\bibfnamefont {A.}~\bibnamefont {Dunsworth}}, \bibinfo {author}
		{\bibfnamefont {D.}~\bibnamefont {Sank}}, \bibinfo {author} {\bibfnamefont
			{C.}~\bibnamefont {Quintana}}, \bibinfo {author} {\bibfnamefont
			{M.}~\bibnamefont {McEwen}}, \bibinfo {author} {\bibfnamefont
			{R.}~\bibnamefont {Barends}}, \bibinfo {author} {\bibfnamefont {P.~V.}\
			\bibnamefont {Klimov}}, \bibinfo {author} {\bibfnamefont {S.}~\bibnamefont
			{Hong}}, \bibinfo {author} {\bibfnamefont {C.}~\bibnamefont {Jones}},
		\bibinfo {author} {\bibfnamefont {A.}~\bibnamefont {Petukhov}}, \bibinfo
		{author} {\bibfnamefont {D.}~\bibnamefont {Kafri}}, \bibinfo {author}
		{\bibfnamefont {S.}~\bibnamefont {Demura}}, \bibinfo {author} {\bibfnamefont
			{B.}~\bibnamefont {Burkett}}, \bibinfo {author} {\bibfnamefont
			{C.}~\bibnamefont {Gidney}}, \bibinfo {author} {\bibfnamefont {A.~G.}\
			\bibnamefont {Fowler}}, \bibinfo {author} {\bibfnamefont {A.}~\bibnamefont
			{Paler}}, \bibinfo {author} {\bibfnamefont {H.}~\bibnamefont {Putterman}},
		\bibinfo {author} {\bibfnamefont {I.}~\bibnamefont {Aleiner}}, \bibinfo
		{author} {\bibfnamefont {F.}~\bibnamefont {Arute}}, \bibinfo {author}
		{\bibfnamefont {K.}~\bibnamefont {Arya}}, \bibinfo {author} {\bibfnamefont
			{R.}~\bibnamefont {Babbush}}, \bibinfo {author} {\bibfnamefont {J.~C.}\
			\bibnamefont {Bardin}}, \bibinfo {author} {\bibfnamefont {A.}~\bibnamefont
			{Bengtsson}}, \bibinfo {author} {\bibfnamefont {A.}~\bibnamefont {Bourassa}},
		\bibinfo {author} {\bibfnamefont {M.}~\bibnamefont {Broughton}}, \bibinfo
		{author} {\bibfnamefont {B.~B.}\ \bibnamefont {Buckley}}, \bibinfo {author}
		{\bibfnamefont {D.~A.}\ \bibnamefont {Buell}}, \bibinfo {author}
		{\bibfnamefont {N.}~\bibnamefont {Bushnell}}, \bibinfo {author}
		{\bibfnamefont {B.}~\bibnamefont {Chiaro}}, \bibinfo {author} {\bibfnamefont
			{R.}~\bibnamefont {Collins}}, \bibinfo {author} {\bibfnamefont
			{W.}~\bibnamefont {Courtney}}, \bibinfo {author} {\bibfnamefont {A.~R.}\
			\bibnamefont {Derk}}, \bibinfo {author} {\bibfnamefont {D.}~\bibnamefont
			{Eppens}}, \bibinfo {author} {\bibfnamefont {C.}~\bibnamefont {Erickson}},
		\bibinfo {author} {\bibfnamefont {E.}~\bibnamefont {Farhi}}, \bibinfo
		{author} {\bibfnamefont {B.}~\bibnamefont {Foxen}}, \bibinfo {author}
		{\bibfnamefont {M.}~\bibnamefont {Giustina}}, \bibinfo {author}
		{\bibfnamefont {A.}~\bibnamefont {Greene}}, \bibinfo {author} {\bibfnamefont
			{J.~A.}\ \bibnamefont {Gross}}, \bibinfo {author} {\bibfnamefont {M.~P.}\
			\bibnamefont {Harrigan}}, \bibinfo {author} {\bibfnamefont {S.~D.}\
			\bibnamefont {Harrington}}, \bibinfo {author} {\bibfnamefont
			{J.}~\bibnamefont {Hilton}}, \bibinfo {author} {\bibfnamefont
			{A.}~\bibnamefont {Ho}}, \bibinfo {author} {\bibfnamefont {T.}~\bibnamefont
			{Huang}}, \bibinfo {author} {\bibfnamefont {W.~J.}\ \bibnamefont {Huggins}},
		\bibinfo {author} {\bibfnamefont {L.~B.}\ \bibnamefont {Ioffe}}, \bibinfo
		{author} {\bibfnamefont {S.~V.}\ \bibnamefont {Isakov}}, \bibinfo {author}
		{\bibfnamefont {E.}~\bibnamefont {Jeffrey}}, \bibinfo {author} {\bibfnamefont
			{Z.}~\bibnamefont {Jiang}}, \bibinfo {author} {\bibfnamefont
			{K.}~\bibnamefont {Kechedzhi}}, \bibinfo {author} {\bibfnamefont
			{S.}~\bibnamefont {Kim}}, \bibinfo {author} {\bibfnamefont {A.}~\bibnamefont
			{Kitaev}}, \bibinfo {author} {\bibfnamefont {F.}~\bibnamefont {Kostritsa}},
		\bibinfo {author} {\bibfnamefont {D.}~\bibnamefont {Landhuis}}, \bibinfo
		{author} {\bibfnamefont {P.}~\bibnamefont {Laptev}}, \bibinfo {author}
		{\bibfnamefont {E.}~\bibnamefont {Lucero}}, \bibinfo {author} {\bibfnamefont
			{O.}~\bibnamefont {Martin}}, \bibinfo {author} {\bibfnamefont {J.~R.}\
			\bibnamefont {McClean}}, \bibinfo {author} {\bibfnamefont {T.}~\bibnamefont
			{McCourt}}, \bibinfo {author} {\bibfnamefont {X.}~\bibnamefont {Mi}},
		\bibinfo {author} {\bibfnamefont {K.~C.}\ \bibnamefont {Miao}}, \bibinfo
		{author} {\bibfnamefont {M.}~\bibnamefont {Mohseni}}, \bibinfo {author}
		{\bibfnamefont {S.}~\bibnamefont {Montazeri}}, \bibinfo {author}
		{\bibfnamefont {W.}~\bibnamefont {Mruczkiewicz}}, \bibinfo {author}
		{\bibfnamefont {J.}~\bibnamefont {Mutus}}, \bibinfo {author} {\bibfnamefont
			{O.}~\bibnamefont {Naaman}}, \bibinfo {author} {\bibfnamefont
			{M.}~\bibnamefont {Neeley}}, \bibinfo {author} {\bibfnamefont
			{C.}~\bibnamefont {Neill}}, \bibinfo {author} {\bibfnamefont
			{M.}~\bibnamefont {Newman}}, \bibinfo {author} {\bibfnamefont {M.~Y.}\
			\bibnamefont {Niu}}, \bibinfo {author} {\bibfnamefont {T.~E.}\ \bibnamefont
			{O'Brien}}, \bibinfo {author} {\bibfnamefont {A.}~\bibnamefont {Opremcak}},
		\bibinfo {author} {\bibfnamefont {E.}~\bibnamefont {Ostby}}, \bibinfo
		{author} {\bibfnamefont {B.}~\bibnamefont {Pat{\'o}}}, \bibinfo {author}
		{\bibfnamefont {N.}~\bibnamefont {Redd}}, \bibinfo {author} {\bibfnamefont
			{P.}~\bibnamefont {Roushan}}, \bibinfo {author} {\bibfnamefont {N.~C.}\
			\bibnamefont {Rubin}}, \bibinfo {author} {\bibfnamefont {V.}~\bibnamefont
			{Shvarts}}, \bibinfo {author} {\bibfnamefont {D.}~\bibnamefont {Strain}},
		\bibinfo {author} {\bibfnamefont {M.}~\bibnamefont {Szalay}}, \bibinfo
		{author} {\bibfnamefont {M.~D.}\ \bibnamefont {Trevithick}}, \bibinfo
		{author} {\bibfnamefont {B.}~\bibnamefont {Villalonga}}, \bibinfo {author}
		{\bibfnamefont {T.}~\bibnamefont {White}}, \bibinfo {author} {\bibfnamefont
			{Z.~J.}\ \bibnamefont {Yao}}, \bibinfo {author} {\bibfnamefont
			{P.}~\bibnamefont {Yeh}}, \bibinfo {author} {\bibfnamefont {J.}~\bibnamefont
			{Yoo}}, \bibinfo {author} {\bibfnamefont {A.}~\bibnamefont {Zalcman}},
		\bibinfo {author} {\bibfnamefont {H.}~\bibnamefont {Neven}}, \bibinfo
		{author} {\bibfnamefont {S.}~\bibnamefont {Boixo}}, \bibinfo {author}
		{\bibfnamefont {V.}~\bibnamefont {Smelyanskiy}}, \bibinfo {author}
		{\bibfnamefont {Y.}~\bibnamefont {Chen}}, \bibinfo {author} {\bibfnamefont
			{A.}~\bibnamefont {Megrant}}, \bibinfo {author} {\bibfnamefont
			{J.}~\bibnamefont {Kelly}},\ and\ \bibinfo {author} {\bibnamefont {{Google
					Quantum AI}}},\ }\bibfield  {title} {\bibinfo {title} {Exponential
			suppression of bit or phase errors with cyclic error correction},\ }\href
	{https://doi.org/10.1038/s41586-021-03588-y} {\bibfield  {journal} {\bibinfo
			{journal} {Nature}\ }\textbf {\bibinfo {volume} {595}},\ \bibinfo {pages}
		{383} (\bibinfo {year} {2021}{\natexlab{a}})}\BibitemShut {NoStop}%
	\bibitem [{\citenamefont {Chen}\ \emph
		{et~al.}(2021{\natexlab{b}})\citenamefont {Chen}, \citenamefont {Yoder},
		\citenamefont {Kim}, \citenamefont {Sundaresan}, \citenamefont {Srinivasan},
		\citenamefont {Li}, \citenamefont {C{\'o}rcoles}, \citenamefont {Cross},\
		and\ \citenamefont {Takita}}]{chen2021calibrated}%
	\BibitemOpen
	\bibfield  {author} {\bibinfo {author} {\bibfnamefont {E.~H.}\ \bibnamefont
			{Chen}}, \bibinfo {author} {\bibfnamefont {T.~J.}\ \bibnamefont {Yoder}},
		\bibinfo {author} {\bibfnamefont {Y.}~\bibnamefont {Kim}}, \bibinfo {author}
		{\bibfnamefont {N.}~\bibnamefont {Sundaresan}}, \bibinfo {author}
		{\bibfnamefont {S.}~\bibnamefont {Srinivasan}}, \bibinfo {author}
		{\bibfnamefont {M.}~\bibnamefont {Li}}, \bibinfo {author} {\bibfnamefont
			{A.~D.}\ \bibnamefont {C{\'o}rcoles}}, \bibinfo {author} {\bibfnamefont
			{A.~W.}\ \bibnamefont {Cross}},\ and\ \bibinfo {author} {\bibfnamefont
			{M.}~\bibnamefont {Takita}},\ }\bibfield  {title} {\bibinfo {title}
		{Calibrated decoders for experimental quantum error correction},\ }\href@noop
	{} {\bibfield  {journal} {\bibinfo  {journal} {arXiv preprint
				arXiv:2110.04285}\ } (\bibinfo {year} {2021}{\natexlab{b}})}\BibitemShut
	{NoStop}%
	\bibitem [{\citenamefont {Marques}\ \emph {et~al.}(2021)\citenamefont
		{Marques}, \citenamefont {Varbanov}, \citenamefont {Moreira}, \citenamefont
		{Ali}, \citenamefont {Muthusubramanian}, \citenamefont {Zachariadis},
		\citenamefont {Battistel}, \citenamefont {Beekman}, \citenamefont {Haider},
		\citenamefont {Vlothuizen}, \citenamefont {Bruno}, \citenamefont {Terhal},\
		and\ \citenamefont
		{DiCarlo}}]{marquesLogicalqubitOperationsErrordetecting2021}%
	\BibitemOpen
	\bibfield  {author} {\bibinfo {author} {\bibfnamefont {J.~F.}\ \bibnamefont
			{Marques}}, \bibinfo {author} {\bibfnamefont {B.~M.}\ \bibnamefont
			{Varbanov}}, \bibinfo {author} {\bibfnamefont {M.~S.}\ \bibnamefont
			{Moreira}}, \bibinfo {author} {\bibfnamefont {H.}~\bibnamefont {Ali}},
		\bibinfo {author} {\bibfnamefont {N.}~\bibnamefont {Muthusubramanian}},
		\bibinfo {author} {\bibfnamefont {C.}~\bibnamefont {Zachariadis}}, \bibinfo
		{author} {\bibfnamefont {F.}~\bibnamefont {Battistel}}, \bibinfo {author}
		{\bibfnamefont {M.}~\bibnamefont {Beekman}}, \bibinfo {author} {\bibfnamefont
			{N.}~\bibnamefont {Haider}}, \bibinfo {author} {\bibfnamefont
			{W.}~\bibnamefont {Vlothuizen}}, \bibinfo {author} {\bibfnamefont
			{A.}~\bibnamefont {Bruno}}, \bibinfo {author} {\bibfnamefont {B.~M.}\
			\bibnamefont {Terhal}},\ and\ \bibinfo {author} {\bibfnamefont
			{L.}~\bibnamefont {DiCarlo}},\ }\bibfield  {title} {\bibinfo {title}
		{Logical-qubit operations in an error-detecting surface code},\ }\href@noop
	{} {\bibfield  {journal} {\bibinfo  {journal} {arXiv:2102.13071 [cond-mat,
				physics:quant-ph]}\ } (\bibinfo {year} {2021})},\ \Eprint
	{https://arxiv.org/abs/2102.13071} {arXiv:2102.13071 [cond-mat,
		physics:quant-ph]} \BibitemShut {NoStop}%
	\bibitem [{\citenamefont {Postler}\ \emph {et~al.}(2021)\citenamefont
		{Postler}, \citenamefont {Heußen}, \citenamefont {Pogorelov}, \citenamefont
		{Rispler}, \citenamefont {Feldker}, \citenamefont {Meth}, \citenamefont
		{Marciniak}, \citenamefont {Stricker}, \citenamefont {Ringbauer},
		\citenamefont {Blatt}, \citenamefont {Schindler}, \citenamefont {Müller},\
		and\ \citenamefont {Monz}}]{postler2021demonstration}%
	\BibitemOpen
	\bibfield  {author} {\bibinfo {author} {\bibfnamefont {L.}~\bibnamefont
			{Postler}}, \bibinfo {author} {\bibfnamefont {S.}~\bibnamefont {Heußen}},
		\bibinfo {author} {\bibfnamefont {I.}~\bibnamefont {Pogorelov}}, \bibinfo
		{author} {\bibfnamefont {M.}~\bibnamefont {Rispler}}, \bibinfo {author}
		{\bibfnamefont {T.}~\bibnamefont {Feldker}}, \bibinfo {author} {\bibfnamefont
			{M.}~\bibnamefont {Meth}}, \bibinfo {author} {\bibfnamefont {C.~D.}\
			\bibnamefont {Marciniak}}, \bibinfo {author} {\bibfnamefont {R.}~\bibnamefont
			{Stricker}}, \bibinfo {author} {\bibfnamefont {M.}~\bibnamefont {Ringbauer}},
		\bibinfo {author} {\bibfnamefont {R.}~\bibnamefont {Blatt}}, \bibinfo
		{author} {\bibfnamefont {P.}~\bibnamefont {Schindler}}, \bibinfo {author}
		{\bibfnamefont {M.}~\bibnamefont {Müller}},\ and\ \bibinfo {author}
		{\bibfnamefont {T.}~\bibnamefont {Monz}},\ }\bibfield  {title} {\bibinfo
		{title} {Demonstration of fault-tolerant universal quantum gate operations},\
	}\href@noop {} {\  (\bibinfo {year} {2021})},\ \Eprint
	{https://arxiv.org/abs/2111.12654} {arXiv:2111.12654 [quant-ph]} \BibitemShut
	{NoStop}%
	\bibitem [{\citenamefont {Krinner}\ \emph {et~al.}(2021)\citenamefont
		{Krinner}, \citenamefont {Lacroix}, \citenamefont {Remm}, \citenamefont
		{Paolo}, \citenamefont {Genois}, \citenamefont {Leroux}, \citenamefont
		{Hellings}, \citenamefont {Lazar}, \citenamefont {Swiadek}, \citenamefont
		{Herrmann}, \citenamefont {Norris}, \citenamefont {Andersen}, \citenamefont
		{Muller}, \citenamefont {Blais}, \citenamefont {Eichler},\ and\ \citenamefont
		{Wallraff}}]{Krinner2021RealizingRQ}%
	\BibitemOpen
	\bibfield  {author} {\bibinfo {author} {\bibfnamefont {S.}~\bibnamefont
			{Krinner}}, \bibinfo {author} {\bibfnamefont {N.}~\bibnamefont {Lacroix}},
		\bibinfo {author} {\bibfnamefont {A.}~\bibnamefont {Remm}}, \bibinfo {author}
		{\bibfnamefont {A.~D.}\ \bibnamefont {Paolo}}, \bibinfo {author}
		{\bibfnamefont {{\'E}.}~\bibnamefont {Genois}}, \bibinfo {author}
		{\bibfnamefont {C.}~\bibnamefont {Leroux}}, \bibinfo {author} {\bibfnamefont
			{C.}~\bibnamefont {Hellings}}, \bibinfo {author} {\bibfnamefont
			{S.}~\bibnamefont {Lazar}}, \bibinfo {author} {\bibfnamefont
			{F.}~\bibnamefont {Swiadek}}, \bibinfo {author} {\bibfnamefont
			{J.}~\bibnamefont {Herrmann}}, \bibinfo {author} {\bibfnamefont {G.~J.}\
			\bibnamefont {Norris}}, \bibinfo {author} {\bibfnamefont {C.~K.}\
			\bibnamefont {Andersen}}, \bibinfo {author} {\bibfnamefont {M.}~\bibnamefont
			{Muller}}, \bibinfo {author} {\bibfnamefont {A.}~\bibnamefont {Blais}},
		\bibinfo {author} {\bibfnamefont {C.}~\bibnamefont {Eichler}},\ and\ \bibinfo
		{author} {\bibfnamefont {A.}~\bibnamefont {Wallraff}},\ }\bibfield  {title}
	{\bibinfo {title} {Realizing repeated quantum error correction in a
			distance-three surface code}\ }(\bibinfo {year} {2021})\BibitemShut {NoStop}%
	\bibitem [{\citenamefont {Zhao}\ \emph {et~al.}(2021)\citenamefont {Zhao},
		\citenamefont {Ye}, \citenamefont {Huang}, \citenamefont {Zhang},
		\citenamefont {Wu}, \citenamefont {Guan}, \citenamefont {Zhu}, \citenamefont
		{Wei}, \citenamefont {He}, \citenamefont {Cao}, \citenamefont {Chen},
		\citenamefont {Chung}, \citenamefont {Deng}, \citenamefont {Fan},
		\citenamefont {Gong}, \citenamefont {Guo}, \citenamefont {Guo}, \citenamefont
		{Han}, \citenamefont {Li}, \citenamefont {Li}, \citenamefont {Li},
		\citenamefont {Liang}, \citenamefont {Lin}, \citenamefont {Qian},
		\citenamefont {Rong}, \citenamefont {Su}, \citenamefont {Sun}, \citenamefont
		{Wang}, \citenamefont {Wu}, \citenamefont {Xu}, \citenamefont {Ying},
		\citenamefont {Yu}, \citenamefont {Zha}, \citenamefont {Zhang}, \citenamefont
		{Huo}, \citenamefont {Lu}, \citenamefont {Peng}, \citenamefont {Zhu},\ and\
		\citenamefont {Pan}}]{Zhao2021RealizingAE}%
	\BibitemOpen
	\bibfield  {author} {\bibinfo {author} {\bibfnamefont {Y.-W.}\ \bibnamefont
			{Zhao}}, \bibinfo {author} {\bibfnamefont {Y.}~\bibnamefont {Ye}}, \bibinfo
		{author} {\bibfnamefont {H.-L.}\ \bibnamefont {Huang}}, \bibinfo {author}
		{\bibfnamefont {Y.}~\bibnamefont {Zhang}}, \bibinfo {author} {\bibfnamefont
			{D.}~\bibnamefont {Wu}}, \bibinfo {author} {\bibfnamefont {H.-R.}\
			\bibnamefont {Guan}}, \bibinfo {author} {\bibfnamefont {Q.}~\bibnamefont
			{Zhu}}, \bibinfo {author} {\bibfnamefont {Z.}~\bibnamefont {Wei}}, \bibinfo
		{author} {\bibfnamefont {T.}~\bibnamefont {He}}, \bibinfo {author}
		{\bibfnamefont {S.}~\bibnamefont {Cao}}, \bibinfo {author} {\bibfnamefont
			{F.}~\bibnamefont {Chen}}, \bibinfo {author} {\bibfnamefont {T.~H.}\
			\bibnamefont {Chung}}, \bibinfo {author} {\bibfnamefont {H.}~\bibnamefont
			{Deng}}, \bibinfo {author} {\bibfnamefont {D.}~\bibnamefont {Fan}}, \bibinfo
		{author} {\bibfnamefont {M.}~\bibnamefont {Gong}}, \bibinfo {author}
		{\bibfnamefont {C.}~\bibnamefont {Guo}}, \bibinfo {author} {\bibfnamefont
			{S.}~\bibnamefont {Guo}}, \bibinfo {author} {\bibfnamefont {L.}~\bibnamefont
			{Han}}, \bibinfo {author} {\bibfnamefont {N.}~\bibnamefont {Li}}, \bibinfo
		{author} {\bibfnamefont {S.}~\bibnamefont {Li}}, \bibinfo {author}
		{\bibfnamefont {Y.}~\bibnamefont {Li}}, \bibinfo {author} {\bibfnamefont
			{F.}~\bibnamefont {Liang}}, \bibinfo {author} {\bibfnamefont
			{J.}~\bibnamefont {Lin}}, \bibinfo {author} {\bibfnamefont {H.}~\bibnamefont
			{Qian}}, \bibinfo {author} {\bibfnamefont {H.}~\bibnamefont {Rong}}, \bibinfo
		{author} {\bibfnamefont {H.}~\bibnamefont {Su}}, \bibinfo {author}
		{\bibfnamefont {L.}~\bibnamefont {Sun}}, \bibinfo {author} {\bibfnamefont
			{S.}~\bibnamefont {Wang}}, \bibinfo {author} {\bibfnamefont {Y.}~\bibnamefont
			{Wu}}, \bibinfo {author} {\bibfnamefont {Y.}~\bibnamefont {Xu}}, \bibinfo
		{author} {\bibfnamefont {C.}~\bibnamefont {Ying}}, \bibinfo {author}
		{\bibfnamefont {J.}~\bibnamefont {Yu}}, \bibinfo {author} {\bibfnamefont
			{C.}~\bibnamefont {Zha}}, \bibinfo {author} {\bibfnamefont {K.}~\bibnamefont
			{Zhang}}, \bibinfo {author} {\bibfnamefont {Y.}~\bibnamefont {Huo}}, \bibinfo
		{author} {\bibfnamefont {C.}~\bibnamefont {Lu}}, \bibinfo {author}
		{\bibfnamefont {C.-Z.}\ \bibnamefont {Peng}}, \bibinfo {author}
		{\bibfnamefont {X.}~\bibnamefont {Zhu}},\ and\ \bibinfo {author}
		{\bibfnamefont {J.-W.}\ \bibnamefont {Pan}},\ }\bibfield  {title} {\bibinfo
		{title} {Realizing an error-correcting surface code with superconducting
			qubits}\ }(\bibinfo {year} {2021})\BibitemShut {NoStop}%
	\bibitem [{\citenamefont {{Ryan-Anderson}}\ \emph {et~al.}(2021)\citenamefont
		{{Ryan-Anderson}}, \citenamefont {Bohnet}, \citenamefont {Lee}, \citenamefont
		{Gresh}, \citenamefont {Hankin}, \citenamefont {Gaebler}, \citenamefont
		{Francois}, \citenamefont {Chernoguzov}, \citenamefont {Lucchetti},
		\citenamefont {Brown}, \citenamefont {Gatterman}, \citenamefont {Halit},
		\citenamefont {Gilmore}, \citenamefont {Gerber}, \citenamefont {Neyenhuis},
		\citenamefont {Hayes},\ and\ \citenamefont
		{Stutz}}]{ryan-andersonRealizationRealtimeFaulttolerant2021}%
	\BibitemOpen
	\bibfield  {author} {\bibinfo {author} {\bibfnamefont {C.}~\bibnamefont
			{{Ryan-Anderson}}}, \bibinfo {author} {\bibfnamefont {J.~G.}\ \bibnamefont
			{Bohnet}}, \bibinfo {author} {\bibfnamefont {K.}~\bibnamefont {Lee}},
		\bibinfo {author} {\bibfnamefont {D.}~\bibnamefont {Gresh}}, \bibinfo
		{author} {\bibfnamefont {A.}~\bibnamefont {Hankin}}, \bibinfo {author}
		{\bibfnamefont {J.~P.}\ \bibnamefont {Gaebler}}, \bibinfo {author}
		{\bibfnamefont {D.}~\bibnamefont {Francois}}, \bibinfo {author}
		{\bibfnamefont {A.}~\bibnamefont {Chernoguzov}}, \bibinfo {author}
		{\bibfnamefont {D.}~\bibnamefont {Lucchetti}}, \bibinfo {author}
		{\bibfnamefont {N.~C.}\ \bibnamefont {Brown}}, \bibinfo {author}
		{\bibfnamefont {T.~M.}\ \bibnamefont {Gatterman}}, \bibinfo {author}
		{\bibfnamefont {S.~K.}\ \bibnamefont {Halit}}, \bibinfo {author}
		{\bibfnamefont {K.}~\bibnamefont {Gilmore}}, \bibinfo {author} {\bibfnamefont
			{J.}~\bibnamefont {Gerber}}, \bibinfo {author} {\bibfnamefont
			{B.}~\bibnamefont {Neyenhuis}}, \bibinfo {author} {\bibfnamefont
			{D.}~\bibnamefont {Hayes}},\ and\ \bibinfo {author} {\bibfnamefont {R.~P.}\
			\bibnamefont {Stutz}},\ }\bibfield  {title} {\bibinfo {title} {Realization of
			real-time fault-tolerant quantum error correction},\ }\href@noop {}
	{\bibfield  {journal} {\bibinfo  {journal} {arXiv:2107.07505 [quant-ph]}\ }
		(\bibinfo {year} {2021})},\ \Eprint {https://arxiv.org/abs/2107.07505}
	{arXiv:2107.07505 [quant-ph]} \BibitemShut {NoStop}%
	\bibitem [{\citenamefont {Egan}\ \emph {et~al.}(2021)\citenamefont {Egan},
		\citenamefont {Debroy}, \citenamefont {Noel}, \citenamefont {Risinger},
		\citenamefont {Zhu}, \citenamefont {Biswas}, \citenamefont {Newman},
		\citenamefont {Li}, \citenamefont {Brown}, \citenamefont {Cetina},\ and\
		\citenamefont {Monroe}}]{eganFaulttolerantControlErrorcorrected2021b}%
	\BibitemOpen
	\bibfield  {author} {\bibinfo {author} {\bibfnamefont {L.}~\bibnamefont
			{Egan}}, \bibinfo {author} {\bibfnamefont {D.~M.}\ \bibnamefont {Debroy}},
		\bibinfo {author} {\bibfnamefont {C.}~\bibnamefont {Noel}}, \bibinfo {author}
		{\bibfnamefont {A.}~\bibnamefont {Risinger}}, \bibinfo {author}
		{\bibfnamefont {D.}~\bibnamefont {Zhu}}, \bibinfo {author} {\bibfnamefont
			{D.}~\bibnamefont {Biswas}}, \bibinfo {author} {\bibfnamefont
			{M.}~\bibnamefont {Newman}}, \bibinfo {author} {\bibfnamefont
			{M.}~\bibnamefont {Li}}, \bibinfo {author} {\bibfnamefont {K.~R.}\
			\bibnamefont {Brown}}, \bibinfo {author} {\bibfnamefont {M.}~\bibnamefont
			{Cetina}},\ and\ \bibinfo {author} {\bibfnamefont {C.}~\bibnamefont
			{Monroe}},\ }\bibfield  {title} {\bibinfo {title} {Fault-tolerant control of
			an error-corrected qubit},\ }\href
	{https://doi.org/10.1038/s41586-021-03928-y} {\bibfield  {journal} {\bibinfo
			{journal} {Nature}\ ,\ \bibinfo {pages} {1}} (\bibinfo {year}
		{2021})}\BibitemShut {NoStop}%
	\bibitem [{\citenamefont {Bacon}(2006)}]{Bacon2006OperatorQE}%
	\BibitemOpen
	\bibfield  {author} {\bibinfo {author} {\bibfnamefont {D.}~\bibnamefont
			{Bacon}},\ }\bibfield  {title} {\bibinfo {title} {Operator quantum
			error-correcting subsystems for self-correcting quantum memories},\
	}\href@noop {} {\bibfield  {journal} {\bibinfo  {journal} {Physical Review
				A}\ }\textbf {\bibinfo {volume} {73}},\ \bibinfo {pages} {012340} (\bibinfo
		{year} {2006})}\BibitemShut {NoStop}%
	\bibitem [{\citenamefont {Pryadko}(2020)}]{pryadko2020maximum}%
	\BibitemOpen
	\bibfield  {author} {\bibinfo {author} {\bibfnamefont {L.~P.}\ \bibnamefont
			{Pryadko}},\ }\bibfield  {title} {\bibinfo {title} {On maximum-likelihood
			decoding with circuit-level errors},\ }\href@noop {} {\bibfield  {journal}
		{\bibinfo  {journal} {Quantum}\ }\textbf {\bibinfo {volume} {4}},\ \bibinfo
		{pages} {304} (\bibinfo {year} {2020})}\BibitemShut {NoStop}%
	\bibitem [{\citenamefont {Bravyi}\ \emph {et~al.}(2014)\citenamefont {Bravyi},
		\citenamefont {Suchara},\ and\ \citenamefont
		{Vargo}}]{bravyiEfficientAlgorithmsMaximum2014}%
	\BibitemOpen
	\bibfield  {author} {\bibinfo {author} {\bibfnamefont {S.}~\bibnamefont
			{Bravyi}}, \bibinfo {author} {\bibfnamefont {M.}~\bibnamefont {Suchara}},\
		and\ \bibinfo {author} {\bibfnamefont {A.}~\bibnamefont {Vargo}},\ }\bibfield
	{title} {\bibinfo {title} {Efficient {{Algorithms}} for {{Maximum Likelihood
					Decoding}} in the {{Surface Code}}},\ }\href
	{https://doi.org/10.1103/PhysRevA.90.032326} {\bibfield  {journal} {\bibinfo
			{journal} {Physical Review A}\ }\textbf {\bibinfo {volume} {90}},\ \bibinfo
		{pages} {032326} (\bibinfo {year} {2014})},\ \Eprint
	{https://arxiv.org/abs/1405.4883} {arXiv:1405.4883} \BibitemShut {NoStop}%
	\bibitem [{\citenamefont {Gottesman}(1998)}]{gottesman1998heisenberg}%
	\BibitemOpen
	\bibfield  {author} {\bibinfo {author} {\bibfnamefont {D.}~\bibnamefont
			{Gottesman}},\ }\bibfield  {title} {\bibinfo {title} {The heisenberg
			representation of quantum computers},\ }\href@noop {} {\bibfield  {journal}
		{\bibinfo  {journal} {arXiv preprint quant-ph/9807006}\ } (\bibinfo {year}
		{1998})}\BibitemShut {NoStop}%
	\bibitem [{Note1()}]{Note1}%
	\BibitemOpen
	\bibinfo {note} {In the distance 2 case of \cite {chen2021calibrated}, the
		hyperedges that did not have a unique logical label were called \protect
		\emph {ambiguous}, and were the motivation for a partial post-selection
		scheme.}\BibitemShut {Stop}%
	\bibitem [{Note2()}]{Note2}%
	\BibitemOpen
	\bibinfo {note} {Flag measurements from qubits 16, 18, 21, 23 are simply
		ignored with no corrections applied. If flag 11 is non-trivial and 12
		trivial, apply $Z$ to 2. If 12 is non-trivial and 11 trivial, apply $Z$ to
		qubit 6. If flag 13 is non-trivial and 14 trivial, apply $Z$ to qubit 7. If
		14 is non-trivial and 13 trivial, apply $Z$ to qubit 8. See \cite
		{chen2021calibrated} for details on why this is sufficient for
		fault-tolerance.}\BibitemShut {Stop}%
	\bibitem [{\citenamefont {Fowler}\ \emph {et~al.}()\citenamefont {Fowler},
		\citenamefont {Whiteside},\ and\ \citenamefont
		{Hollenberg}}]{Fowler2012matching}%
	\BibitemOpen
	\bibfield  {author} {\bibinfo {author} {\bibfnamefont {A.~G.}\ \bibnamefont
			{Fowler}}, \bibinfo {author} {\bibfnamefont {A.~C.}\ \bibnamefont
			{Whiteside}},\ and\ \bibinfo {author} {\bibfnamefont {L.~C.~L.}\ \bibnamefont
			{Hollenberg}},\ }\bibfield  {title} {\bibinfo {title} {Towards practical
			classical processing for the surface code},\ }\href@noop {} {\bibinfo
		{journal} {Phys. Rev. Lett.}\ }\BibitemShut {NoStop}%
	\bibitem [{\citenamefont {Higgott}(2021)}]{higgottPyMatchingPythonPackage2021}%
	\BibitemOpen
	\bibfield  {journal} {  }\bibfield  {author} {\bibinfo {author} {\bibfnamefont
			{O.}~\bibnamefont {Higgott}},\ }\bibfield  {title} {\bibinfo {title}
		{{{PyMatching}}: A {{Python}} package for decoding quantum codes with
			minimum-weight perfect matching},\ }\href@noop {} {\bibfield  {journal}
		{\bibinfo  {journal} {arXiv:2105.13082 [quant-ph]}\ } (\bibinfo {year}
		{2021})},\ \Eprint {https://arxiv.org/abs/2105.13082} {arXiv:2105.13082
		[quant-ph]} \BibitemShut {NoStop}%
	\bibitem [{\citenamefont {Dua}\ \emph {et~al.}(2022)\citenamefont {Dua},
		\citenamefont {Jochym-O'Connor},\ and\ \citenamefont {Zhu}}]{Dua2022fractal}%
	\BibitemOpen
	\bibfield  {author} {\bibinfo {author} {\bibfnamefont {A.}~\bibnamefont
			{Dua}}, \bibinfo {author} {\bibfnamefont {T.}~\bibnamefont
			{Jochym-O'Connor}},\ and\ \bibinfo {author} {\bibfnamefont {G.}~\bibnamefont
			{Zhu}},\ }\bibfield  {title} {\bibinfo {title} {Quantum error correction with
			fractal topological codes},\ }\href@noop {} {\bibfield  {journal} {\bibinfo
			{journal} {arXiv:2201.03568}\ } (\bibinfo {year} {2022})}\BibitemShut
	{NoStop}%
	\bibitem [{\citenamefont {Bravyi}\ and\ \citenamefont
		{Cross}(2015)}]{bravyiDoubledColorCodes2015}%
	\BibitemOpen
	\bibfield  {author} {\bibinfo {author} {\bibfnamefont {S.}~\bibnamefont
			{Bravyi}}\ and\ \bibinfo {author} {\bibfnamefont {A.}~\bibnamefont {Cross}},\
	}\bibfield  {title} {\bibinfo {title} {Doubled {{Color Codes}}},\ }\href@noop
	{} {\bibfield  {journal} {\bibinfo  {journal} {arXiv:1509.03239 [quant-ph]}\
		} (\bibinfo {year} {2015})},\ \Eprint {https://arxiv.org/abs/1509.03239}
	{arXiv:1509.03239 [quant-ph]} \BibitemShut {NoStop}%
	\bibitem [{\citenamefont {Heim}\ \emph {et~al.}(2016)\citenamefont {Heim},
		\citenamefont {Svore},\ and\ \citenamefont
		{Hastings}}]{heimOptimalCircuitLevelDecoding2016}%
	\BibitemOpen
	\bibfield  {author} {\bibinfo {author} {\bibfnamefont {B.}~\bibnamefont
			{Heim}}, \bibinfo {author} {\bibfnamefont {K.~M.}\ \bibnamefont {Svore}},\
		and\ \bibinfo {author} {\bibfnamefont {M.~B.}\ \bibnamefont {Hastings}},\
	}\bibfield  {title} {\bibinfo {title} {Optimal {{Circuit}}-{{Level Decoding}}
			for {{Surface Codes}}},\ }\href@noop {} {\bibfield  {journal} {\bibinfo
			{journal} {arXiv:1609.06373 [quant-ph]}\ } (\bibinfo {year} {2016})},\
	\Eprint {https://arxiv.org/abs/1609.06373} {arXiv:1609.06373 [quant-ph]}
	\BibitemShut {NoStop}%
	\bibitem [{Note3()}]{Note3}%
	\BibitemOpen
	\bibinfo {note} {$ibm\protect \_peekskill$ v2.0.0, an IBM Quantum Falcon R8
		processor. https://quantum-computing.ibm.com/, accessed Jan.
		2022}\BibitemShut {NoStop}%
	\bibitem [{\citenamefont {{IBM Quantum}}\ and\ \citenamefont
		{{Community}}(2021)}]{Qiskit}%
	\BibitemOpen
	\bibfield  {author} {\bibinfo {author} {\bibnamefont {{IBM Quantum}}}\ and\
		\bibinfo {author} {\bibnamefont {{Community}}},\ }\href
	{https://doi.org/10.5281/zenodo.2573505} {\bibinfo {title} {Qiskit: An
			open-source framework for quantum computing}} (\bibinfo {year}
	{2021})\BibitemShut {NoStop}%
	\bibitem [{\citenamefont {McKay}\ \emph {et~al.}(2019)\citenamefont {McKay},
		\citenamefont {Sheldon}, \citenamefont {Smolin}, \citenamefont {Chow},\ and\
		\citenamefont {Gambetta}}]{McKay19_3QRB}%
	\BibitemOpen
	\bibfield  {author} {\bibinfo {author} {\bibfnamefont {D.~C.}\ \bibnamefont
			{McKay}}, \bibinfo {author} {\bibfnamefont {S.}~\bibnamefont {Sheldon}},
		\bibinfo {author} {\bibfnamefont {J.~A.}\ \bibnamefont {Smolin}}, \bibinfo
		{author} {\bibfnamefont {J.~M.}\ \bibnamefont {Chow}},\ and\ \bibinfo
		{author} {\bibfnamefont {J.~M.}\ \bibnamefont {Gambetta}},\ }\bibfield
	{title} {\bibinfo {title} {Three-qubit randomized benchmarking},\ }\href
	{https://doi.org/10.1103/PhysRevLett.122.200502} {\bibfield  {journal}
		{\bibinfo  {journal} {Phys. Rev. Lett.}\ }\textbf {\bibinfo {volume} {122}},\
		\bibinfo {pages} {200502} (\bibinfo {year} {2019})}\BibitemShut {NoStop}%
	\bibitem [{\citenamefont {Sundaresan}\ \emph {et~al.}(2020)\citenamefont
		{Sundaresan}, \citenamefont {Lauer}, \citenamefont {Pritchett}, \citenamefont
		{Magesan}, \citenamefont {Jurcevic},\ and\ \citenamefont
		{Gambetta}}]{sundaresanReducingUnitarySpectator2020}%
	\BibitemOpen
	\bibfield  {author} {\bibinfo {author} {\bibfnamefont {N.}~\bibnamefont
			{Sundaresan}}, \bibinfo {author} {\bibfnamefont {I.}~\bibnamefont {Lauer}},
		\bibinfo {author} {\bibfnamefont {E.}~\bibnamefont {Pritchett}}, \bibinfo
		{author} {\bibfnamefont {E.}~\bibnamefont {Magesan}}, \bibinfo {author}
		{\bibfnamefont {P.}~\bibnamefont {Jurcevic}},\ and\ \bibinfo {author}
		{\bibfnamefont {J.~M.}\ \bibnamefont {Gambetta}},\ }\bibfield  {title}
	{\bibinfo {title} {Reducing {{Unitary}} and {{Spectator Errors}} in {{Cross
					Resonance}} with {{Optimized Rotary Echoes}}},\ }\href
	{https://doi.org/10.1103/PRXQuantum.1.020318} {\bibfield  {journal} {\bibinfo
			{journal} {PRX Quantum}\ }\textbf {\bibinfo {volume} {1}},\ \bibinfo {pages}
		{020318} (\bibinfo {year} {2020})}\BibitemShut {NoStop}%
	\bibitem [{\citenamefont {Byrd}\ \emph {et~al.}(1995)\citenamefont {Byrd},
		\citenamefont {Lu}, \citenamefont {Nocedal},\ and\ \citenamefont
		{Zhu}}]{byrd1995limited}%
	\BibitemOpen
	\bibfield  {author} {\bibinfo {author} {\bibfnamefont {R.~H.}\ \bibnamefont
			{Byrd}}, \bibinfo {author} {\bibfnamefont {P.}~\bibnamefont {Lu}}, \bibinfo
		{author} {\bibfnamefont {J.}~\bibnamefont {Nocedal}},\ and\ \bibinfo {author}
		{\bibfnamefont {C.}~\bibnamefont {Zhu}},\ }\bibfield  {title} {\bibinfo
		{title} {A limited memory algorithm for bound constrained optimization},\
	}\href@noop {} {\bibfield  {journal} {\bibinfo  {journal} {SIAM Journal on
				scientific computing}\ }\textbf {\bibinfo {volume} {16}},\ \bibinfo {pages}
		{1190} (\bibinfo {year} {1995})}\BibitemShut {NoStop}%
	\bibitem [{\citenamefont {Wood}\ and\ \citenamefont
		{Gambetta}(2018)}]{Wood2018Quantification}%
	\BibitemOpen
	\bibfield  {author} {\bibinfo {author} {\bibfnamefont {C.~J.}\ \bibnamefont
			{Wood}}\ and\ \bibinfo {author} {\bibfnamefont {J.~M.}\ \bibnamefont
			{Gambetta}},\ }\bibfield  {title} {\bibinfo {title} {Quantification and
			characterization of leakage errors},\ }\href
	{https://doi.org/10.1103/PhysRevA.97.032306} {\bibfield  {journal} {\bibinfo
			{journal} {Phys. Rev. A}\ }\textbf {\bibinfo {volume} {97}},\ \bibinfo
		{pages} {032306} (\bibinfo {year} {2018})}\BibitemShut {NoStop}%
	\bibitem [{\citenamefont {Sank}\ \emph {et~al.}(2016)\citenamefont {Sank},
		\citenamefont {Chen}, \citenamefont {Khezri}, \citenamefont {Kelly},
		\citenamefont {Barends}, \citenamefont {Campbell}, \citenamefont {Chen},
		\citenamefont {Chiaro}, \citenamefont {Dunsworth}, \citenamefont {Fowler},
		\citenamefont {Jeffrey}, \citenamefont {Lucero}, \citenamefont {Megrant},
		\citenamefont {Mutus}, \citenamefont {Neeley}, \citenamefont {Neill},
		\citenamefont {O'Malley}, \citenamefont {Quintana}, \citenamefont {Roushan},
		\citenamefont {Vainsencher}, \citenamefont {White}, \citenamefont {Wenner},
		\citenamefont {Korotkov},\ and\ \citenamefont
		{Martinis}}]{Sank2016Measurement}%
	\BibitemOpen
	\bibfield  {author} {\bibinfo {author} {\bibfnamefont {D.}~\bibnamefont
			{Sank}}, \bibinfo {author} {\bibfnamefont {Z.}~\bibnamefont {Chen}}, \bibinfo
		{author} {\bibfnamefont {M.}~\bibnamefont {Khezri}}, \bibinfo {author}
		{\bibfnamefont {J.}~\bibnamefont {Kelly}}, \bibinfo {author} {\bibfnamefont
			{R.}~\bibnamefont {Barends}}, \bibinfo {author} {\bibfnamefont
			{B.}~\bibnamefont {Campbell}}, \bibinfo {author} {\bibfnamefont
			{Y.}~\bibnamefont {Chen}}, \bibinfo {author} {\bibfnamefont {B.}~\bibnamefont
			{Chiaro}}, \bibinfo {author} {\bibfnamefont {A.}~\bibnamefont {Dunsworth}},
		\bibinfo {author} {\bibfnamefont {A.}~\bibnamefont {Fowler}}, \bibinfo
		{author} {\bibfnamefont {E.}~\bibnamefont {Jeffrey}}, \bibinfo {author}
		{\bibfnamefont {E.}~\bibnamefont {Lucero}}, \bibinfo {author} {\bibfnamefont
			{A.}~\bibnamefont {Megrant}}, \bibinfo {author} {\bibfnamefont
			{J.}~\bibnamefont {Mutus}}, \bibinfo {author} {\bibfnamefont
			{M.}~\bibnamefont {Neeley}}, \bibinfo {author} {\bibfnamefont
			{C.}~\bibnamefont {Neill}}, \bibinfo {author} {\bibfnamefont {P.~J.~J.}\
			\bibnamefont {O'Malley}}, \bibinfo {author} {\bibfnamefont {C.}~\bibnamefont
			{Quintana}}, \bibinfo {author} {\bibfnamefont {P.}~\bibnamefont {Roushan}},
		\bibinfo {author} {\bibfnamefont {A.}~\bibnamefont {Vainsencher}}, \bibinfo
		{author} {\bibfnamefont {T.}~\bibnamefont {White}}, \bibinfo {author}
		{\bibfnamefont {J.}~\bibnamefont {Wenner}}, \bibinfo {author} {\bibfnamefont
			{A.~N.}\ \bibnamefont {Korotkov}},\ and\ \bibinfo {author} {\bibfnamefont
			{J.~M.}\ \bibnamefont {Martinis}},\ }\bibfield  {title} {\bibinfo {title}
		{Measurement-induced state transitions in a superconducting qubit: Beyond the
			rotating wave approximation},\ }\href
	{https://doi.org/10.1103/PhysRevLett.117.190503} {\bibfield  {journal}
		{\bibinfo  {journal} {Phys. Rev. Lett.}\ }\textbf {\bibinfo {volume} {117}},\
		\bibinfo {pages} {190503} (\bibinfo {year} {2016})}\BibitemShut {NoStop}%
	\bibitem [{\citenamefont {Bravyi}\ \emph {et~al.}(2021)\citenamefont {Bravyi},
		\citenamefont {Sheldon}, \citenamefont {Kandala}, \citenamefont {Mckay},\
		and\ \citenamefont {Gambetta}}]{Bravyi2021Mitigating}%
	\BibitemOpen
	\bibfield  {author} {\bibinfo {author} {\bibfnamefont {S.}~\bibnamefont
			{Bravyi}}, \bibinfo {author} {\bibfnamefont {S.}~\bibnamefont {Sheldon}},
		\bibinfo {author} {\bibfnamefont {A.}~\bibnamefont {Kandala}}, \bibinfo
		{author} {\bibfnamefont {D.~C.}\ \bibnamefont {Mckay}},\ and\ \bibinfo
		{author} {\bibfnamefont {J.~M.}\ \bibnamefont {Gambetta}},\ }\bibfield
	{title} {\bibinfo {title} {Mitigating measurement errors in multiqubit
			experiments},\ }\href {https://doi.org/10.1103/PhysRevA.103.042605}
	{\bibfield  {journal} {\bibinfo  {journal} {Phys. Rev. A}\ }\textbf {\bibinfo
			{volume} {103}},\ \bibinfo {pages} {042605} (\bibinfo {year}
		{2021})}\BibitemShut {NoStop}%
	\bibitem [{\citenamefont {Lescanne}\ \emph {et~al.}(2019)\citenamefont
		{Lescanne}, \citenamefont {Verney}, \citenamefont {Ficheux}, \citenamefont
		{Devoret}, \citenamefont {Huard}, \citenamefont {Mirrahimi},\ and\
		\citenamefont {Leghtas}}]{Lescanne19}%
	\BibitemOpen
	\bibfield  {author} {\bibinfo {author} {\bibfnamefont {R.}~\bibnamefont
			{Lescanne}}, \bibinfo {author} {\bibfnamefont {L.}~\bibnamefont {Verney}},
		\bibinfo {author} {\bibfnamefont {Q.}~\bibnamefont {Ficheux}}, \bibinfo
		{author} {\bibfnamefont {M.~H.}\ \bibnamefont {Devoret}}, \bibinfo {author}
		{\bibfnamefont {B.}~\bibnamefont {Huard}}, \bibinfo {author} {\bibfnamefont
			{M.}~\bibnamefont {Mirrahimi}},\ and\ \bibinfo {author} {\bibfnamefont
			{Z.}~\bibnamefont {Leghtas}},\ }\bibfield  {title} {\bibinfo {title} {Escape
			of a driven quantum josephson circuit into unconfined states},\ }\href
	{https://doi.org/10.1103/PhysRevApplied.11.014030} {\bibfield  {journal}
		{\bibinfo  {journal} {Phys. Rev. Applied}\ }\textbf {\bibinfo {volume}
			{11}},\ \bibinfo {pages} {014030} (\bibinfo {year} {2019})}\BibitemShut
	{NoStop}%
	\bibitem [{\citenamefont {McEwen}\ \emph {et~al.}(2021)\citenamefont {McEwen},
		\citenamefont {Kafri}, \citenamefont {Chen}, \citenamefont {Atalaya},
		\citenamefont {Satzinger}, \citenamefont {Quintana}, \citenamefont {Klimov},
		\citenamefont {Sank}, \citenamefont {Gidney}, \citenamefont {Fowler},
		\citenamefont {Arute}, \citenamefont {Arya}, \citenamefont {Buckley},
		\citenamefont {Burkett}, \citenamefont {Bushnell}, \citenamefont {Chiaro},
		\citenamefont {Collins}, \citenamefont {Demura}, \citenamefont {Dunsworth},
		\citenamefont {Erickson}, \citenamefont {Foxen}, \citenamefont {Giustina},
		\citenamefont {Huang}, \citenamefont {Hong}, \citenamefont {Jeffrey},
		\citenamefont {Kim}, \citenamefont {Kechedzhi}, \citenamefont {Kostritsa},
		\citenamefont {Laptev}, \citenamefont {Megrant}, \citenamefont {Mi},
		\citenamefont {Mutus}, \citenamefont {Naaman}, \citenamefont {Neeley},
		\citenamefont {Neill}, \citenamefont {Niu}, \citenamefont {Paler},
		\citenamefont {Redd}, \citenamefont {Roushan}, \citenamefont {White},
		\citenamefont {Yao}, \citenamefont {Yeh}, \citenamefont {Zalcman},
		\citenamefont {Chen}, \citenamefont {Smelyanskiy}, \citenamefont {Martinis},
		\citenamefont {Neven}, \citenamefont {Kelly}, \citenamefont {Korotkov},
		\citenamefont {Petukhov},\ and\ \citenamefont {Barends}}]{McEwen21}%
	\BibitemOpen
	\bibfield  {author} {\bibinfo {author} {\bibfnamefont {M.}~\bibnamefont
			{McEwen}}, \bibinfo {author} {\bibfnamefont {D.}~\bibnamefont {Kafri}},
		\bibinfo {author} {\bibfnamefont {Z.}~\bibnamefont {Chen}}, \bibinfo {author}
		{\bibfnamefont {J.}~\bibnamefont {Atalaya}}, \bibinfo {author} {\bibfnamefont
			{K.~J.}\ \bibnamefont {Satzinger}}, \bibinfo {author} {\bibfnamefont
			{C.}~\bibnamefont {Quintana}}, \bibinfo {author} {\bibfnamefont {P.~V.}\
			\bibnamefont {Klimov}}, \bibinfo {author} {\bibfnamefont {D.}~\bibnamefont
			{Sank}}, \bibinfo {author} {\bibfnamefont {C.}~\bibnamefont {Gidney}},
		\bibinfo {author} {\bibfnamefont {A.~G.}\ \bibnamefont {Fowler}}, \bibinfo
		{author} {\bibfnamefont {F.}~\bibnamefont {Arute}}, \bibinfo {author}
		{\bibfnamefont {K.}~\bibnamefont {Arya}}, \bibinfo {author} {\bibfnamefont
			{B.}~\bibnamefont {Buckley}}, \bibinfo {author} {\bibfnamefont
			{B.}~\bibnamefont {Burkett}}, \bibinfo {author} {\bibfnamefont
			{N.}~\bibnamefont {Bushnell}}, \bibinfo {author} {\bibfnamefont
			{B.}~\bibnamefont {Chiaro}}, \bibinfo {author} {\bibfnamefont
			{R.}~\bibnamefont {Collins}}, \bibinfo {author} {\bibfnamefont
			{S.}~\bibnamefont {Demura}}, \bibinfo {author} {\bibfnamefont
			{A.}~\bibnamefont {Dunsworth}}, \bibinfo {author} {\bibfnamefont
			{C.}~\bibnamefont {Erickson}}, \bibinfo {author} {\bibfnamefont
			{B.}~\bibnamefont {Foxen}}, \bibinfo {author} {\bibfnamefont
			{M.}~\bibnamefont {Giustina}}, \bibinfo {author} {\bibfnamefont
			{T.}~\bibnamefont {Huang}}, \bibinfo {author} {\bibfnamefont
			{S.}~\bibnamefont {Hong}}, \bibinfo {author} {\bibfnamefont {E.}~\bibnamefont
			{Jeffrey}}, \bibinfo {author} {\bibfnamefont {S.}~\bibnamefont {Kim}},
		\bibinfo {author} {\bibfnamefont {K.}~\bibnamefont {Kechedzhi}}, \bibinfo
		{author} {\bibfnamefont {F.}~\bibnamefont {Kostritsa}}, \bibinfo {author}
		{\bibfnamefont {P.}~\bibnamefont {Laptev}}, \bibinfo {author} {\bibfnamefont
			{A.}~\bibnamefont {Megrant}}, \bibinfo {author} {\bibfnamefont
			{X.}~\bibnamefont {Mi}}, \bibinfo {author} {\bibfnamefont {J.}~\bibnamefont
			{Mutus}}, \bibinfo {author} {\bibfnamefont {O.}~\bibnamefont {Naaman}},
		\bibinfo {author} {\bibfnamefont {M.}~\bibnamefont {Neeley}}, \bibinfo
		{author} {\bibfnamefont {C.}~\bibnamefont {Neill}}, \bibinfo {author}
		{\bibfnamefont {M.}~\bibnamefont {Niu}}, \bibinfo {author} {\bibfnamefont
			{A.}~\bibnamefont {Paler}}, \bibinfo {author} {\bibfnamefont
			{N.}~\bibnamefont {Redd}}, \bibinfo {author} {\bibfnamefont {P.}~\bibnamefont
			{Roushan}}, \bibinfo {author} {\bibfnamefont {T.~C.}\ \bibnamefont {White}},
		\bibinfo {author} {\bibfnamefont {J.}~\bibnamefont {Yao}}, \bibinfo {author}
		{\bibfnamefont {P.}~\bibnamefont {Yeh}}, \bibinfo {author} {\bibfnamefont
			{A.}~\bibnamefont {Zalcman}}, \bibinfo {author} {\bibfnamefont
			{Y.}~\bibnamefont {Chen}}, \bibinfo {author} {\bibfnamefont {V.~N.}\
			\bibnamefont {Smelyanskiy}}, \bibinfo {author} {\bibfnamefont {J.~M.}\
			\bibnamefont {Martinis}}, \bibinfo {author} {\bibfnamefont {H.}~\bibnamefont
			{Neven}}, \bibinfo {author} {\bibfnamefont {J.}~\bibnamefont {Kelly}},
		\bibinfo {author} {\bibfnamefont {A.~N.}\ \bibnamefont {Korotkov}}, \bibinfo
		{author} {\bibfnamefont {A.~G.}\ \bibnamefont {Petukhov}},\ and\ \bibinfo
		{author} {\bibfnamefont {R.}~\bibnamefont {Barends}},\ }\bibfield  {title}
	{\bibinfo {title} {Removing leakage-induced correlated errors in
			superconducting quantum error correction},\ }\href@noop {} {\bibfield
		{journal} {\bibinfo  {journal} {Nature Communications}\ }\textbf {\bibinfo
			{volume} {12}},\ \bibinfo {pages} {1761} (\bibinfo {year}
		{2021})}\BibitemShut {NoStop}%
	\bibitem [{\citenamefont {Battistel}\ \emph {et~al.}(2021)\citenamefont
		{Battistel}, \citenamefont {Varbanov},\ and\ \citenamefont
		{Terhal}}]{Battistel21}%
	\BibitemOpen
	\bibfield  {author} {\bibinfo {author} {\bibfnamefont {F.}~\bibnamefont
			{Battistel}}, \bibinfo {author} {\bibfnamefont {B.}~\bibnamefont
			{Varbanov}},\ and\ \bibinfo {author} {\bibfnamefont {B.}~\bibnamefont
			{Terhal}},\ }\bibfield  {title} {\bibinfo {title} {Hardware-efficient
			leakage-reduction scheme for quantum error correction with superconducting
			transmon qubits},\ }\href {https://doi.org/10.1103/PRXQuantum.2.030314}
	{\bibfield  {journal} {\bibinfo  {journal} {PRX Quantum}\ }\textbf {\bibinfo
			{volume} {2}},\ \bibinfo {pages} {030314} (\bibinfo {year}
		{2021})}\BibitemShut {NoStop}%
	\bibitem [{\citenamefont {Suchara}\ \emph {et~al.}(2014)\citenamefont
		{Suchara}, \citenamefont {Cross},\ and\ \citenamefont
		{Gambetta}}]{SucharaLRU}%
	\BibitemOpen
	\bibfield  {author} {\bibinfo {author} {\bibfnamefont {M.}~\bibnamefont
			{Suchara}}, \bibinfo {author} {\bibfnamefont {A.~W.}\ \bibnamefont {Cross}},\
		and\ \bibinfo {author} {\bibfnamefont {J.~M.}\ \bibnamefont {Gambetta}},\
	}\bibfield  {title} {\bibinfo {title} {Leakage suppression in the toric
			code},\ }\href@noop {} {\bibfield  {journal} {\bibinfo  {journal}
			{arXiv:1410.8562}\ } (\bibinfo {year} {2014})}\BibitemShut {NoStop}%
\end{thebibliography}
\end{document}